\documentclass[a4paper,11pt]{article}

\usepackage{hyperref}

\usepackage[utf8]{inputenc}
\usepackage[english, spanish]{babel}
\usepackage{graphicx}
\usepackage{epsfig}
\usepackage{amssymb}
\usepackage{bm}
\usepackage{amsmath}
\usepackage{color}
\usepackage{hyperref}
\usepackage{times}
\usepackage{float}
\usepackage{times}

\def\be{\begin{equation}}
\def\te{\end{equation}}
\def\ee{\end{equation}}
\def\ba{\begin{eqnarray}}
\def\bea{\begin{eqnarray}}
\def\nn{\nonumber\\}
\def\tea{\end{eqnarray}}
\def\ea{\end{eqnarray}}
\def\eea{\end{eqnarray}}
\begin{document}

\title{Propagation Speeds of Relativistic Conformal Particles from a
Generalized Relaxation Time Approximation}

\author{Alejandra Kandus\\
LATO-DCET Universidade Estadual de Santa Cruz,Ilhéus-BA, Brazil\thanks{kandus@uesc.br}\\[0.5cm]
Esteban Calzetta\\
Universidad de Buenos Aires, Facultad de Ciencias Exactas y Naturales,\\ Departamento de Física, Argentina,\\ y CONICET - Universidad de Buenos Aires,\\ Instituto de Física de Buenos Aires (IFIBA), Argentina\thanks{calzetta@df.uba.ar}}

\maketitle

\begin{abstract}{The propagation speeds of excitations are a crucial input in the modeling of interacting systems of particles. In this paper we assume the microscopic physics is described by a kinetic theory for massless particles, which is approximated by a generalized relaxation time approximation (RTA) where the relaxation time depends on the energy of the particles involved. We seek a solution of the kinetic equation by assuming a parameterized one-particle distribution function (1-pdf) which generalizes the Chapman-Enskog (Ch-En) solution to the RTA. If developed to all orders, this would yield an asymptotic solution to the kinetic equation; we restrict ourselves to an approximate solution by truncating the Ch-En series to second order. Our generalized Ch-En solution contains undetermined space-time dependent parameters, and we derive a set of dynamical equations for them applying the moments method. We check that these dynamical equations lead to energy-momentum conservation and positive entropy production. Finally, we compute the propagation speeds for fluctuations way from equilibrium from the linearized form of the dynamical equations. Considering relaxation times of the form $\tau=\tau_0(-\beta_{\mu}p^{\mu})^{-a}$, with $-\infty< a<2$, where $\beta_{\mu}=u_{\mu}/T$ is the temperature vector in the Landau frame, we show that the Anderson-Witting prescription $a=1$ yields the fastest speed in all scalar, vector and tensor sectors. This fact ought to be taken into consideration when choosing the best macroscopic description for a given physical system.}
\end{abstract}

\section{Introduction}\label{intro}

The success of hydrodynamics in the description of the early stages of relativistic heavy ion collisions \cite{Rom-Rom-19,calzetta16} and the promise of relevant cosmological applications 
\cite{GraCalKan22,GraCal18,MiGra21} have turned the study of strongly interacting systems of relativistic particles into an active area of research \cite{tanos,DRlibro,CalHu08}. Although kinetic theory provides the microscopic description for these systems \cite{Isr76,LMR86,GL90,PRCal09}, the full Boltzmann equations are generally regarded as too complex, and simpler schemes consistent with the most important physical features are sought.  Among these simpler schemes, the Relaxation Time Approximation (RTA), which assumes a collision term parameterized by a relaxation time, stands out. The RTA includes the non linearities in the theory through the equilibrium one particle distribution function (1-pdf) towards which the system relaxes, and in this sense the RTA ansatz contains the footprint of the non-linear and 
complete underlying kinetic theory that describes the microphysics.

{The first implementations of the Bhatnagar, Gross and Krook
BGK RTA \cite{BGK54} to relativistic fluids are those of Marle 
\cite{Marle-a,Marle-b} and Anderson and Witting (AW) \cite{AWa,AWb}. Both expressions are phenomenological ansätze, proportional to $(f-f_0)$ and in each case $f_0$ corresponds to different 'frames' (see sub-section \ref{Ma-AW} below). In the case of Marle, the proportionality factor is $m/\tau$, with $m$ the mass of the gas particles, while in that of AW it is $u_{\mu}p^{\mu}/\tau$. In both cases the parameter $\tau$ is identified with a 'relaxation time'. In principle the choice of the proportionality factor must be guided by phenomenological considerations or else by a systematic derivation form the Boltzmann equation, which is feasible in some cases. See Refs. \cite{DeNo24,RBD23b}. However successful, both Marle and AW RTA's severely distort the structure of the linearized Boltzmann equation, and their validity is doubtful for \emph{soft} collision terms, where the continuous spectrum of the linearized collision operator reaches up to the zero eigenvalue associated with the hydrodynamic modes. \cite{Dud13,LY16,Hu24}}.

This has led to more general implementations of the RTA \cite{DMT10,LO10} where different modes were allowed to relax at different rates. However, simply allowing the relaxation times to be a function of energy other than constant or linear is not satisfactory as it may violate energy-momentum conservation \cite{RDN21,RFDN22}. Rather the RTA must be implemented preserving the Hilbert space structure of the space of linearized one-particle distribution function (1-pdf's). A kinetic equation allowing for a  momentum dependent relaxation time consistent with energy-momentum conservation was  introduced in \cite{PRCal10,PRCal13}. In this paper we shall elaborate on this proposal. In particular, we shall show how to produce a RTA matching any prescribed spectrum for the linearized collision operator, either soft or hard. Not been able to do this is one of the main drawbacks of the usual formulations of the RTA. See also \cite{KW19,WW21,M21,M22,AM23,AM24,H23,DBJJ22,DBJJ23,B23}

Although a general solution of the RTA kinetic equation may be attempted, this leads to an integral equation for the equilibrium 1-pdf, which must be related to the actual 1-pdf through some prescription, which will be different for different choices of the relaxation time. For example, under the Marle prescription the equilibrium 1-pdf leads to the same particle current than the actual one, while under the Anderson-Witting prescription, the equilibrium and actual 1-pdfs are matched though the energy current. This obscures the physical features of the system, which close enough to equilibrium is dominated by the hydrodynamic modes and a few long-lived non-hydrodynamic modes. To capture this behavior, it is best to assume a parameterized form for the 1-pdf, supplemented by dynamical equations for the parameters. In this paper, we shall obtain these dynamical equations by taking moments of the kinetic equation. 

Of course this poses the challenge of finding a suitable parameterization for the 1-pdf. The parameterization should be general enough to allow for an accurate description of physically meaningful processes but not so general as to make the ensuing theory unwieldly. Many proposals have been advanced in the literature \cite{RWDNR24,S24,Gav24a,WG24,GaDiNo24a,GaDiNo24b} .

In this paper we shall adopt the point of view that the parameterization must be such as to include the Chapman-Enskog (Ch-En) solution to the kinetic equation as a particular case \cite{ChapCow}. In other words, we shall use the second order Ch-En as a template, generalizing it so to obtain a family of parameterized 1-pdf, still containing the actual Ch-En solution as a particular case.

{One reason for working this way is the fact that in the Ch-En expansion, as in  DNMR 
\cite{DNMR12}  and IReD \cite{WPA22}, each order is determined by the power of a certain small parameter. Thus it is easy to identify the intensity of the deviation from the local thermal balance.}

The procedure we propose yields parameterizations with an increasing complexity depending on to which order the Ch-En solution is computed. In this manuscript we shall work to second order \cite{BKCJ20,BKCJ21,DMMEZ23}, which in the AW case returns the theory already analyzed in \cite{PeCa21}.

Once the parameterization has been chosen, the next step is to find equations of motion for the parameters. We require that these equations both conserve energy-momentum and enforce the Second Law. Note that even if the kinetic theory allows for a $H$-theorem, a positive entropy production in the parameterized theory does not follow automatically,  because the parameterized 1-pdf is not a solution of the kinetic equation. As shown in  \cite{CanCal20}, a suitable set of equations of motion is derived by taking the moments of the kinetic equation against the same irreducible phase-space functions which appear in the parameterization, see below, eq. (\ref{moments}). Observe that in principle this method does not require linearization on deviations from equilibrium (see also \cite{AgCal17}), though in general the resulting theory is too complex unless severely restricted by symmetry considerations.

The ultimate goal of this paper is to compute the propagation speeds for collective modes of a conformal real relativistic system of interacting particles \cite{S05,BR79,BR97a,BR97b,BR99,Mull99}. We define the propagation speed as the velocity of a front across which the parameters are continuous but their first derivatives are not.
To derive the hydrodynamic equations we apply the moments method to a kinetic equation under the RTA. 
We work in the 
Landau frame of the fluid, and generalize the usual RTA by allowing an arbitrary dependence of relaxation time on the particle energy  \cite{PRCal10,PRCal13}. Different choices for this dependence mean different underlying
kinetic theories or equivalently, microphysics. 

{One crucial step in this procedure is the choice of which moments of the 1-pdf shall be considered. As we said above, we adopt the criterium of choosing the functions of momentum which make up a second order Chapman-Enskog solution. In other words, we work within a restricted class of 1-pdfs which generalizes the second order Chapman-Enskog solution, retaining it as a particular case. This choice leads us to parametrize the 1-pdf in terms of functions of momentum which themselves depend on the way the relaxation time relates to energy, see eqs. (\ref{xlist}) below. Because of this, different choices of the functional dependence of the relaxation time lead to different dynamical equations for the system, not only in the terms which derive from the moments of the collision integral, where the dependence on the relaxation time is explicit, but also in the derivative terms. }
We thus obtain a family of dynamical theories with different phenomenology
according to the choice of the  relaxation time as a function of energy.
We could understand this dependence as the remaining footprint of a non-linear and complete kinetic theory, which was approximated by a mathematically more tractable RTA.

In summary, assuming the relaxation time dependent upon energy \cite{RDN21,RFDN22,PRCal10,PRCal13} we show, on one hand, how to compare the predictions from different choices of this dependence, and, on the other, that those different choices lead to macroscopic models that are clearly distinguishable (in this case by producing different propagation speeds).

The propagation speeds of a theory are of course fundamental to determine causality. 
The theory we are considering here is thermodinamically stable by construction, and our results will confirm the expectation that it is causal as well \cite{SM22,HSSW23,ASN23,BRD23,HK24,WP24,GDN24}. 

Propagation speeds are also relevant for the discussion of shocks \cite{S05,Isr60,Cal22}. The propagation speed in kinetic theory is the velocity of the fastest particle for which the 1-pdf is not zero, and so it can be arbitrarily close to the speed of light $c$ for a suitable 1-pdf \cite{Isr88}. There are also examples from field theory where the propagation speed is arbitrarily close to that of light \cite{Moore24}. Hydrodynamics, on the other hand, usually has a fastest propagation speed which is less than $c$ by a finite amount \cite{BR99}. For this reason, strong enough shocks in hydrodynamics are discontinuous. This discontinuity is not observed in kinetic theory, and may be regarded as an artifact of the hydrodynamic approximation. When considering approximations to the full kinetic equations, as in this paper, the issue of which setup yields the fastest speed becomes most relevant, as this is also the framework which provides the best description of shocks.

To make the discussion more concrete we shall consider a particular family of generalized RTA's where the relaxation time takes the form $\tau=\tau_0(-\beta_{\mu}p^{\mu})^{-a}$, with $-\infty< a<2$, where $\beta_{\mu}=u_{\mu}/T$ is the temperature vector in the Landau frame of the fluid. This family covers both the case where hard modes thermalize faster than soft modes and the converse. It also contains Marle's and AW RTA's as the $a=0$ and $a=1$ particular cases, respectively. The upper limit in $a$ is necessary to avoid infrared divergences in the equations of motion.

We find that the  AW choice $a=1$, where we recover the results of \cite{PeCa21}, yields the fastest speeds. 

To summarize, the main results of this paper are a) the construction of a generalized RTA designed to match the spectrum of any linearized kinetic equation, b) the derivation thereby of a parameterized theory which is causal and stable and enforces both energy-momentum conservation and positive entropy production, c) the computation of the propagation speeds for scalar, vector and tensor perturbations away from equilibrium for a family of generalized RTA's containing Marle's and AW's as particular cases, and d) the verification that AW RTA yields the fastest speed within this family.

This paper is organized as follows: In Section \ref{KT} we shortly review the features of Kinetic Theory, and in section \ref{RT} we elaborate on the generalized RTA as a substitute for the actual kinetic equation as derived from microphysics. In Section \ref{2oCH} we deduce the 1-pdf Ch-En solution to the Boltzmann equation up to second order in gradients, for a momentum dependent RTA.  We then introduce a 1-pdf after the pattern of the Ch-En solution and derive the  set of moment equations, which, for the purposes of finding the propagation speeds, may be particularized at the free-streaming regime. In Section \ref{ps} we perform a scalar-vector-tensor decomposition and write down the equation system corresponding to each sector. To have a glimpse of their solutions we consider the family of RTA's given by $\tau=\tau_0(-\beta_{\mu}p^{\mu})^{-a}$, with $-\infty < a \leq 2$ to avoid infrared divergences. In Section \ref{conc} we summarize the main conclusions. Details of the derivation of the 
Ch-En solution are given in the appendix \ref{Chensol}.

We work with natural units $\hbar=c=k_B=1$ and signature $\left(-,+,+,+\right)$.

\section{Relativistic Kinetic theory}\label{KT}

The central object of a kinetic description is the one-particle distribution function (1-pdf) which gives the probability of finding a particle within a given phase space cell, at a particular event and with a particular momentum, constrained to be on mass shell and to have positive energy \cite{Isr72,GLW80,Ste71,Lib03}. For simplicity we shall consider only gases whose equilibrium distribution is of the Maxwell-Jüttner kind eq. (\ref{MJ}). The 1-pdf is advected by the particles and changes because of collisions among particles. Therefore the kinetic equation has a transport part and a collision integral which gives the change in the 1-pdf due to collisions per unit particle proper time, as in

\be 
p^{\mu}\frac{\partial f}{\partial x^{\mu}}=\mathcal{I}_{coll}
\label{kinetic}
\te 
There is fairly universal agreement about the transport part, while different kinetic approaches posit different collisions operators \cite{CalHu08}.  A collision integral must be consistent with energy momentum conservation (for simplicity, we shall deal below with a massless particles, and thus we do not impose particle number conservation) and allow for an $H$ theorem, entropy production being zero only for Maxwell-Jüttner 1-pdfs

\be 
f_0=e^{\beta_{\mu}p^{\mu}}\label{MJ}
\te 
where $\beta_{\mu}=u_{\mu}/T$, with $T$ being the fluid temperature and $u_{\mu}$ the velocity in a frame to be chosen below, with $u^2=-1$. Then

\be
\mathcal{I}_{coll}\left[ f_0\right] =0 \label{icoll-f0}
\te 
The energy momentum tensor (EMT) and entropy flux are 

\bea
&&T^{\mu\nu}=\int Dp\;p^{\mu}p^{\nu}f\label{Tab-defined}\\
&&S^{\mu}=\int Dp\;p^{\mu}f\left[ 1-\ln f\right] \label{Sa-defined}
\tea
where 

\be 
Dp=2\frac{d^4p}{\left( 2\pi\right)^3 }\delta\left( -p^2\right) \theta\left( p^0\right) 
\label{med-def}
\te 
is the Lorentz-invariant momentum space volume element. Energy-momentum conservation then implies that

\be 
\int Dp\;p^{\mu}\mathcal{I}_{coll}=0
\label{emtcons}
\te 
and the $H$ theorem

\be 
-\int Dp\;\ln f \;\mathcal{I}_{coll}\ge 0
\label{htheorem}
\te 
for \emph{any} function $f$. \emph{If} $f$ is a solution of the 
kinetic equation eq. (\ref{kinetic}) this leads to positive entropy production $S^{\mu}_{,\mu}\ge 0$.

The Landau-Lifshitz prescription provides a way to associate an inverse temperature vector to any 1-pdf, even if it is not of the Maxwell-Jüttner kind. Namely, we identify the four-velocity with the only timelike eigenvector of the EMT

\be
T^{\mu}_{\nu}u^{\nu}=-\rho u^{\mu} \label{emt-eigenv}
\te 
whereby we identify $\rho$ as the energy density, and then we derive a temperature from $\rho$ by imposing the equilibrium dependence

\be
\rho=\frac 3{\pi^2}T^4 \label{rho-def}
\te
Having identified $\beta^{\mu}$, we may build the corresponding Maxwell-Jüttner 1-pdf $f_0$, eq. (\ref{MJ}). Moreover, the fact that $\ln f$ appears explicitly in the H-theorem eq. (\ref{htheorem}) suggests we decompose $f$ as

\be
f=f_0e^{\chi}
\label{chidec}
\te
Without loss of generality we may also write 

\be
\mathcal{I}_{coll}\left[f\right]=f_0{I}_{coll}\left[\chi\right]
\label{lincoll}
\te 
By virtue of eq. (\ref{emtcons}) $\ln f_0$ does not contribute to entropy creation, and then we find

\be 
S^{\mu}_{;\mu}=-\int Dp\;f_0\;\chi \;{I}_{coll}\left[\chi\right]\ge 0
\label{htheorem2}
\te 
If $f$ is a solution of the kinetic equation.

The ansatz eq. (\ref{chidec}) guarantees a positive one-particle distribution function, even in a full nonlinear theory. Moreover, the entropy production eq. (\ref{htheorem2}) already singles out the logarithm of the one-particle distribution function as playing a most important role. Using this feature we will introduce below an ansatz for the collision term (see eq. (\ref{JPREC} )) which leads to positive entropy production to all orders in deviations from equilibrium, (see eq. (\ref{htheorem3})).

\subsection{Parameterized Kinetic Theory}

Let us look for solutions of eq. (\ref{kinetic}) of the form

\be 
f\equiv f_{hydro}=e^{\sum_{\alpha=0}^nC_{\alpha}X^{\alpha}}
\label{fhydro}
\te
In eq. (\ref{fhydro}), the $C_{\alpha}$ denote tensor fields in space-time, while the $X^{\alpha}$ are tensor fields in phase space. Note that $\alpha$ is not itself a tensorial index, it just numbers the different tensors in the theory. The contractions $C_{\alpha}X^{\alpha}$ are world scalars. In particular, we shall choose
$X^0=p^{\mu}$ and $C_0=\beta_{\mu}$, or equivalently, in the terms of eq. (\ref{chidec})

\be
\chi=\sum_{\alpha=1}^nC_{\alpha}X^{\alpha}
\label{chihydro}
\te
We shall assume $C_{\alpha}\left(x\right)$ are totally symmetric, traceless and transverse tensors, for $\alpha\ge 1$. Of course, if we would allow the $X^{\alpha}$ functions to form a complete set in phase space we could seek an exact solution of the kinetic equations under this form. We shall however truncate the sum in eq. (\ref{chihydro}) at a finite value of $n$, to be specified below, so we shall get only an approximate solution.

The EMT and entropy flux are obtained by substituting $f_{hydro}$ into equations (\ref{Tab-defined}) and (\ref{Sa-defined}). We obtain the four conservation laws (\ref{emtcons}) but these equations are not enough to determine the evolution of the whole set of $C_{\alpha}$ functions, for which we must provide supplementary equations.

Our guiding principle is to obtain positive entropy production. Now

\be
S^{\mu}_{hydro,\mu}=-\int\;Dp\;\left[\sum_{\alpha=1}^nC_{\alpha}X^{\alpha}\right]p^{\mu}f_{hydro,\mu}
\label{2ndLaw-hydro}
\te
The problem is that we cannot call upon the H-theorem to enforce positive entropy production, because $f_{hydro}$ is not a solution of the kinetic equation. We demand instead the moment equations

\be
\int Dp\;X^{\alpha}\left\{p^{\mu}f_{hydro,\mu}-f_0I_{coll}\left[\chi\right]\right\}=0
\label{moments}
\te
which for $\alpha=0$ is just EMT conservation. We now are allowed to substitute

\be
S^{\mu}_{hydro,\mu}=-\int\;Dp\;f_0\chi I_{coll}\left[\chi\right]\ge 0
\label{2ndLaw-hydro2}
\te
The moment equations (\ref{moments}) are thus the equations of motion of the parameterized theory.

This setup enforces positive entropy production but does not tell us how to choose the $X^{\alpha}$ functions, beyond $\alpha=0$. We shall return to this vexing question in section (\ref{2oCH}), after we have introduced the relaxation time approximation.

\section{Relaxation time approximation}\label{RT}

Physically, the role of the collision term in the kinetic equation (\ref{kinetic}) is to force $\chi$ to relax to zero, or at least to a multiple of $p^{\mu}$, the only possibilities leading to vanishing entropy production. A realistic kinetic equation such as Boltzmann's typically leads to a very complex collision term. However, it may be expected that the essentials of the relaxation of $\chi$  may be captured by a much simpler collision term, linear in $\chi$

\be
{I}_{coll}\left[\chi\right]\left(x,p\right)=\int Dp'\;f_0\left(x,p'\right)K\left[p,p'\right]\chi\left(x,p'\right)
\label{grta}
\te
Linearization of the Boltzmann collision term yields a $K$ operator which is symmetric in the space of momentum functions with the inner product \cite{CMK02,Dud13,LY16}

\be
\left\langle \chi'\vert\chi\right\rangle=\int Dp\;f_0\chi'\chi \label{mean-def}
\te 
namely

\be 
\int Dp\;f_0\chi'{I}_{coll}\left[\chi\right]=\int Dp\;f_0\chi{I}_{coll}\left[\chi'\right]
\label{sym-inn-prod}
\te 
${I}_{coll}\left[\chi\right]$ has exactly four  null eigenvectors corresponding to the hydrodynamic modes $\chi_{\mu}=p_{\mu}$; this enforces energy momentum conservation. Moreover the $H$ theorem eq. (\ref{htheorem2}) requires that all non zero eigenvalues of the collision operator must be negative.

We shall call a kinetic equation with a collision term as in eq. (\ref{grta}) a generalized relaxation time approximation. The first relativistic RTA was Marle's \cite{Marle-a,Marle-b}, who wrote the collision operator of the form

\be
I^{\left(M\right)}_{coll}\left[f\right]=\frac{\left(-T\right)}{\tau}\left[f-f^{\left(M\right)}_0\right]
\label{Marle}
\te 
{We must mention that Marle's original expression has a mass $m$ and not a temperature $T$. But as we work with massless particles, the only relevant dimensionful parameter is $T$}. Energy momentum conservation requires

\be
\int Dp\;p^{\mu}\left[f-f^{\left(M\right)}_0\right]=\int Dp\;p^{\mu}f-nu^{\mu}=0
\label{mar-eck}
\te 
where $n=T^3/\pi^2$ would be the particle density for massless particles. Thus the Marle equation requires us to work in the so-called Eckart frame: we identify the velocity and temperature by matching the particle current of the actual 1-pdf \cite{Eck40}. 

After Marle, Anderson and Witting \cite{AWa,AWb} proposed

\be
I^{\left(AW\right)}_{coll}\left[f\right]=\frac{\left(u_{\nu}p^{\nu}\right)}{\tau}\left[f-f^{\left(AW\right)}_0\right]
\label{AWCT}
\te 
so now

\be
\int Dp\;p^{\mu}u_{\nu}p^{\nu}\left[f-f^{\left(AW\right)}_0\right]=u_{\nu}T^{\mu\nu}+\rho u^{\mu}=0 \label{aw-ll}
\te 
where $\rho=3T^4/\pi^2$ is the energy density for a conformal fluid; we see that in the AW formulation $f_0$ is the equilibrium solution in the Landau-Lifshitz frame \cite{LL59}.
{Both Marle's and AW choices seriously distort the Boltzmann dynamics, and are actually disfavoured by experimental data from relativistic heavy ion collisions \cite{DMT10,LO10}}.

\subsection{Generalized Relaxation time approximation}\label{gen-RTA-appr}

Concretely, our concern is to go beyond the Marle and Anderson-Witting RTAs by allowing the relaxation time to depend on the energy of the particle in nontrivial ways. It is clearly seen that trying to improve on Marle's or AW's equations by allowing $\tau$ to be momentum-dependent, while keeping the Eckart or Landau-Lifshitz prescriptions to identify the inverse temperature vector, leads to a contradiction \cite{RDN21,RFDN22}. In this section we shall review the collision term proposed in \cite{PRCal10,PRCal13}, which overcomes this difficulty. For simplicity we shall work in the Landau-Lifshitz frame throughout; this has the appealing feature that it may be determined from the properties of the macroscopic energy-momentum tensor alone.

We introduce the notation

\be
\left\langle X\right\rangle=\left\langle 1\vert X\right\rangle=\int Dp\;f_0\;X
\label{MV-def}
\te
where $f_0$ is the Maxwell-Jüttner distribution eq. (\ref{MJ}) built from the Landau-Lifshitz temperature and velocity. We shall assume the constraint

\be
u_{\mu}\left\langle p^{\mu}p^{\nu}\chi\right\rangle=0
\label{constraint}
\te 
which follows from applying the Landau-Lifshitz prescription to linear order in the deviation from equilibrium $\chi$.

We write the collision integral as in eq. (\ref{lincoll}). Energy momentum conservation requires ${I}_{coll}\left[\chi\right]$ to be orthogonal to the four null eigenvectors $p^{\mu}$. To satisfy this requirement we introduce a projection operator $Q$ such that for any $g$

\be
\left\langle p^{\mu}Q\left[g\right]\right\rangle=0 \label{Q-def}
\te
It is symmetric
\be
\left\langle g'Q\left[g\right]\right\rangle=\left\langle gQ\left[g'\right]\right\rangle
\label{Q-sym}
\te
and 

\be
Q\left[g\right]=g\;\;\;
\iff\;\;\; \left\langle p^{\mu}g\right\rangle=0 \label{Q-g-Q}
\te
These properties suggest that

\be
Q\left[g\right]=g-p^{\nu}{T}_{0\nu\rho}^{-1}\left\langle p^{\rho}g\right\rangle
\label{qdeg}
\te
with

\be
T_0^{\mu\nu}=\left\langle p^{\mu}p^{\nu}\right\rangle=\rho\left[u^{\mu}u^{\nu}+\frac13\Delta^{\mu\nu}\right]
\label{emt-2}
\te
(where $\Delta^{\mu\nu}=\eta^{\mu\nu}+u^{\mu}u^{\nu}$) is the energy-momentum tensor built from $f_0$, and

\be
T_{0\mu\nu}^{-1}=\frac1{\rho}\left[u_{\mu}u_{\nu}+3\Delta_{\mu\nu}\right]
\label{mathcalT}
\te
From $\left\langle p^{\mu}Q\left[g\right]\right\rangle=0$ and the symmetry of $Q$ we conclude that $\left\langle gQ\left[p^{\mu}\right]\right\rangle=0$ and since $g$ is arbitrary it must be $Q\left[p^{\mu}\right]=0$, which is easily verified explicitly. Conversely, if $Q\left[g\right]=0$ then $g=\delta\beta_{\mu}p^{\mu}$ for some momentum independent coefficients $\delta\beta_{\mu}$. Finally, acting with $Q$ on both sides of eq. (\ref{qdeg}) we see that $Q^2=Q$, so $Q$ is indeed a projection.

We may now define the collision integral. To preserve the symmetry we propose

\be
{I}_{coll}\left[\chi\right]=-\frac{T^2}{\varsigma}Q\left[FQ\left[\chi\right]\right]
\label{JPREC}
\te
Where $\varsigma$ is a dimensionless relaxation time and $F=F\left[-\beta_{\mu}p^{\mu}\right]$ is a dimensionless
function. The entropy production eq. (\ref{htheorem2}) becomes

\be
S^{\mu}_{,\mu}=\frac{T^2}{\varsigma}\left\langle F\left(Q\left[\chi\right]\right)^2\right\rangle \label{htheorem3}
\te
Therefore the $H$ theorem requires $F\ge 0$.

{The rationale of the proposal eq. (\ref{JPREC}) is to retain the fundamental features of the Boltzmann equation within a single mathematical structure yet keeping it flexible enough to accommodate phenomenological considerations. Among the former, the feature we want to keep is that the linearized Boltzmann operator is a symmetric operator in a certain Hilbert space \cite{CMK02}, which may be therefore diagonalized, and whose spectrum bears basic information about the physics of the system, most notably whether we deal with a \emph{hard} or \emph{soft} collision term - in the former case the everpresent zero eigenvalue is an isolated eigenvalue, while in the latter it is part of the continuous spectrum \cite{Dud13,LY16,Hu24}. We elaborate on this in the next Subsection \ref{SP}. 
Note that the double projection operator $Q$ in eq. (\ref{JPREC}) both makes the collision term symmetric and enforces energy momentum conservation within a single frame regardless of the function $F$. In this paper we choose to work in the Landau frame throughout, and expect to explore different frame choices in forthcoming work \cite{DoGaMoShTo22,BhaMiRoSi24,BhaMiRo24}.}

On the other hand, we leave a window open for phenomenology through the choice of the function $F$ in eq. (\ref{JPREC}). By far the most common choices for $F$ are Marle's ($F= constant$) and Anderson-Witting's ($F\propto -u_{\mu}p^{\mu}$) to be discussed in more detail below (Section \ref{Ma-AW}). 
However, phenomenological considerations in the context of RHICs have led to the proposal of more general power laws \cite{DMT10,LO10}, which in some cases may be systematically derived from the Boltzmann equation 
\cite{DeNo24,RBD23b}
and are actively under research \cite{Hu24,RDN21,RFDN22,KW19,WW21,M21,M22,AM23,AM24,H23,DBJJ22,DBJJ23,B23}.. In this paper we only consider functions $F$ defined through power laws; even within this restricted class we find propagation speeds are strongly dependent on the function $F$.

\subsection{Spectral considerations}\label{SP}

Let us analyze the equation

\be
Q\left[g\right]=h \label{spectral-1}
\te
We have the integrability conditions $\left\langle p^{\mu}h\right\rangle=0$ (or else, $Q\left[h\right]=Q^2\left[g\right]=Q\left[g\right]=h$), so $h$ itself is a particular solution. Since the $p^{\mu}$'s are homogeneous solutions (note that here $\mu$ is not a world index, it simply distinguishes each of four different functions from each other), the general solution is

\be
g=h+c_{\mu}p^{\mu}
\label{gensol}
\te
We may now analyze the spectrum of ${I}_{coll}$. Suppose 

\be 
Q\left[FQ\left[\zeta_{\lambda}\right]\right]=\lambda\zeta_{\lambda}
\label{spectral-2}
\te
If $\lambda=0$, then $FQ\left[\zeta_{\lambda}\right]=c_{\mu}p^{\mu}$, and then

\be
Q\left[\zeta_{\lambda}\right]=\frac{c_{\mu}p^{\mu}}{F} \label{spectral-3}
\te 
Therefore we must have the integrability condition

\be 
c_{\nu}\left\langle \frac{p^{\mu}p^{\nu}}{F}\right\rangle=0 \label{spectral-4}
\te
but this is impossible unless the $c_{\mu}=0$ themselves. So we must have $Q\left[\zeta_{\lambda}\right]=0$. We conclude that the only null eigenvectors are indeed the $p^{\mu}$ functions.

Now assume $\lambda\not= 0$. Then $Q\left[\zeta_{\lambda}\right]=\zeta_{\lambda}$ and therefore we may write

\be 
Q\left[F\zeta_{\lambda}\right]=\lambda\zeta_{\lambda}
\label{simp}
\te
with general solution (cfr. eq. (\ref{gensol}))

\be
F\zeta_{\lambda}=\lambda\zeta_{\lambda}+c_{\mu}p^{\mu}
\label{gensol2}
\te
If $F\left[p^{\mu}\right]\not=\lambda$ for every $p^{\mu}$, then 

\be
\zeta_{\lambda}=\frac{c_{\mu}p^{\mu}}{F-\lambda} \label{spectral-5}
\te 
but this is not possible because it violates the integrability condition for eq. (\ref{simp}) $\left\langle p^{\mu}\zeta_{\lambda}\right\rangle=0$. Therefore we conclude that the spectrum of the collision operator is included in the image of $F$.

Now assume that $F\left[p^{\mu}\right]=\lambda$ for some $p^{\mu}_{\lambda}$. Let us work in the Landau-Lifshitz rest frame where $u^{\mu}=\left(1,0,0,0\right)$. Then the solution to eq. (\ref{gensol2}) is

\be
\zeta_{\lambda}=f_{\lambda}\left[\frac{\vec p}p\right]\delta\left(F-\lambda\right)+PV\left\{\frac{c_{\mu}p^{\mu}}{F-\lambda}\right\} \label{spectral-6}
\te 
for some function $f_{\lambda}$. We have two possibilities. If 

\be
\left\langle p^{\mu}f_{\lambda}\left[\frac{\vec p}p\right]\delta\left(F-\lambda\right)\right\rangle=0 \label{spectral-7}
\te 
then from $\left\langle p^{\mu}\zeta_{\lambda}\right\rangle=0$ we conclude that the $c_{\mu}$ themselves are zero. Otherwise we get a linear equation from which we determine the $c_{\mu}$ coefficients. Thus
we see that we may easily find a function $F$ to match any preordained spectrum for the collision operator.

\subsection{Marle and Anderson-Witting}\label{Ma-AW}

To conclude this section, we shall discuss whether it is possible to regard the Marle eq. (\ref{Marle}) and AW eq. (\ref{AWCT}) as particular cases of the collision term eq. (\ref{JPREC})

To make contact with AW's equation eq. (\ref{AWCT}) we set $F=F^{\left(AW\right)}\left[x\right]=x$. Recall that since we are defining the velocity and temperature according to the Landau-Lifshitz prescription, when we split the 1-pdf as in eq. (\ref{chidec}), eq. (\ref{constraint}) follows, and we get $\left\langle p^{\mu}F\chi\right\rangle=-\beta_{\nu}\left\langle p^{\mu}p^{\nu}\chi\right\rangle=0$, so $Q\left[F\chi\right]=F\chi$.

Assume we also have $\left\langle p^{\rho}\chi\right\rangle=0$ besides eq. (\ref{constraint}). Then $Q\left[\chi\right]=\chi$, 
and so $Q\left[FQ\left[\chi\right]\right]=Q\left[F\chi\right]=F\chi$, yielding the AW RTA.  In the following we shall refer to $F\left[x\right]=x$ as the Anderson-Witting prescription. 

When we implement Marle's proposal, we must take into account that Marle's fiducial equilibrium 1-pdf is built from matching the particle rather than the energy flux, namely

\be
\int Dp\;p^{\mu}f_0^{\left(M\right)}=\frac{T^{\left(M\right)3}}{\pi^2}u^{\left(M\right)\mu}=\int Dp\;p^{\mu}f=\frac{T^{3}}{\pi^2}u^{\mu}+\left\langle p^{\mu}\chi\right\rangle \label{Ma-AW-1}
\te
Therefore, writing $T^{\left(M\right)}=T\left(1+\delta T^{M}\right)$ and $u^{\left(M\right)\mu}=u^{\mu}+\delta u^{\left(M\right)\mu}$ (with $u_{\mu}\delta u^{\left(M\right)\mu}=0$) we get

\bea
\delta T^{M}&=&-\frac{\pi^2}{3T^3}u_{\mu}\left\langle p^{\mu}\chi\right\rangle\nn
\delta u^{\left(M\right)\mu}&=&\frac{\pi^2}{T^3}\Delta^{\mu}_{\rho}\left\langle p^{\rho}\chi\right\rangle \label{Ma-AW-2}
\tea
It follows that 

\be
f_0^{\left(M\right)}=f_0\left[1+p^{\mu}{T}^{-1}_{0\mu\rho}\left\langle p^{\rho}\chi\right\rangle\right] \label{Ma-AW-3}
\te
From eq. (\ref{qdeg}) we see that the Marle collision integral eq. (\ref{Marle}) is just the collision integral from eq. (\ref{JPREC}) with $F=1$.

In the following section we implement this formalism to obtain a second order theory of relativistic conformal fluids.

\section{Hydrodynamics from the second order Chapman-Enskog solution}\label{2oCH}

As we have mentioned in the Introduction, the first (and probably main) challenge in seeking a parameterized solution to kinetic theory is to find a suitable parameterization of the kinetic 1-pdf. Our proposal is to use the second order Ch-En solution as a template. This means, we shall work out the second order Ch-En solution and then write it as in eq. (\ref{fhydro}), thus identifying the $X^{\alpha}\left(x,p\right)$ functions. Of course, in the actual solution these functions are multiplied by given coefficients built from $\beta^{\mu}$ and its derivatives. We shall later replace these coefficients by unknown functions $C_{\alpha}$ obeying the equations of motion (\ref{moments}), thus obtaining a parameterization that generalizes the Ch-En solution.

The Ch-En solution is a systematic expansion of the 1-pdf in powers of the dimensionless relaxation time $\varsigma$ introduced in eq. (\ref{JPREC}) \cite{ChapCow}. We therefore have a hierarchy of solutions, depending on to which order we extend the expansion.

In the moments or Grad approximation on the other side, there is no explicit small parameter with respect to which we can do a perturbative expansion. Thus it is not clear a priori how many moments of the 1-pdf must be included to describe a given departure from equilibrium. To circumvent this situation, in this paper we take the point of view that the parameterized 1-pdf should take the form eqs. (\ref{fhydro}) and (\ref{chihydro}), where the $X_n$ are the same functions of momentum as they appear in a Ch-En solution at some given order. 

We use a fiducial temperature $T_0$ to build explicit dimensionless quantities, namely, we define $t=T/T_0$ and similarly we make all other quantities non-dimensional by dividing or multiplying by $T_0$ as required. For simplicity we shall not introduce new names for the dimensionless quantities. The dimensionless Boltzmann equation eq. (\ref{JPREC}) reads

\begin{equation}
p^{\mu}\frac{\partial}{\partial x^{\mu}}f = - f_0\frac{{Q}\left[t^2 F\left[-\beta_{\nu} p^{\nu}\right] 
 {Q}\left[\chi\right]\right]}{\varsigma} \label{BolE-1}
\end{equation}
To implement a perturbative scheme we expand
\begin{equation}
 \chi = \sum_{n=1}^{\infty} \varsigma^n \chi_n \label{Ch-E-hydro-1}
\end{equation}
and replacing into eq. (\ref{BolE-1}) we obtain
\be
\varsigma p^{\mu} \left\{\left[ p^{\nu}\beta_{\nu,\mu}
\left(1 + \sum_{n=1}^{\infty} \varsigma^n \chi_n \right)\right]
+ \sum_{n=1}^{\infty} \varsigma^n \chi_{n,\mu}\right\}
=  -\sum_{n=1}^{\infty}\varsigma^{n}  {Q}\left[t^2 F\left[-\beta_{\nu}p^{\nu}\right] 
 {Q}\left[\chi_n\right]\right]
\label{che1}
\te
where we have linearized the transport term, which is accurate enough for the discussion below.

The Ch-En procedure aims at obtaining a solution of this equation as a expansion in powers of $\varsigma$. Space derivatives in the left hand side of eq. (\ref{che1}) are considered to be of ``zeroth order'', while time derivatives, defined as $\dot X=u^{\mu}X_{,\mu}$ have their own development in powers of $\varsigma$

\be 
\dot X=\sum_{n=0}^{\infty}\varsigma^n\dot X^{\left(n\right)} \label{Ch-E-hydro-2}
\te
Replacing in eq. (\ref{che1}) and matching powers of $\varsigma$ we obtain

\bea
&-&p^{\mu}u_{\mu} \left\{p^{\nu}\dot\beta_{\nu}^{\left(n\right)}
+\sum_{m=1}^{n}p^{\nu}\dot\beta_{\nu}^{n-m}\chi_m+\sum_{m=1}^{n}\dot\chi_m^{\left(n-m\right)}\right\}\nn
&+&\delta_{n0}p^{\rho} \Delta^{\mu}_{\rho} p^{\nu}\beta_{\nu,\mu}+p^{\rho} \Delta^{\mu}_{\rho} p^{\nu}\beta_{\nu,\mu}\chi_n+p^{\rho} \Delta^{\mu}_{\rho} \chi_{n,\mu}\nn
&=&  -{Q}\left[t^2 F\left[-\beta_{\nu}p^{\nu}\right]  {Q}\left[\chi_{n+1}\right]\right]
\label{che3}
\tea
Because of the projector in the right hand side at each order we have an integrability condition

\bea
0&=&-u_{\mu} \left\{\left\langle p^{\lambda}p^{\mu}p^{\nu}\right\rangle\dot\beta_{\nu}^{\left(n\right)}+\sum_{m=1}^{n}\left\langle p^{\lambda}p^{\mu}p^{\nu}\chi_m\right\rangle\dot\beta_{\nu}^{n-m}+\sum_{m=1}^{n}\left\langle p^{\lambda}p^{\mu}\dot\chi_m^{\left(n-m\right)}\right\rangle\right\}\nn
&+&\delta_{n0}\Delta^{\mu}_{\rho}\left\langle p^{\lambda}p^{\rho}  p^{\nu}\right\rangle\beta_{\nu,\mu}+\Delta^{\mu}_{\rho}\left\langle p^{\lambda}p^{\rho}  p^{\nu}\chi_n\right\rangle\beta_{\nu,\mu}+ \Delta^{\mu}_{\rho} \left\langle p^{\lambda}p^{\rho} \chi_{n,\mu}\right\rangle
\label{che4}
\tea
We shall simplify these expressions by further linearizing on the $t$ and $u_{\mu}$ derivatives. Then eq. (\ref{che3}) reduces to 

\be
-p^{\mu}u_{\mu} \left\{p^{\nu}\dot\beta_{\nu}^{\left(n\right)}+\sum_{m=1}^{n}\dot\chi_m^{\left(n-m\right)}\right\}+\delta_{n0}p^{\rho} \Delta^{\mu}_{\rho} p^{\nu}\beta_{\nu,\mu}+p^{\rho} \Delta^{\mu}_{\rho} \chi_{n,\mu}= -{Q}\left[t^2 F  {Q}\left[\chi_{n+1}\right]\right]
\label{che3b}
\te
and eq. (\ref{che4}) yields

\be
0=-u_{\mu} \left\{\left\langle p^{\lambda}p^{\mu}p^{\nu}\right\rangle\dot\beta_{\nu}^{\left(n\right)}+\sum_{m=1}^{n}\left\langle p^{\lambda}p^{\mu}\dot\chi_m^{\left(n-m\right)}\right\rangle\right\}+\delta_{n0}\Delta^{\mu}_{\rho}\left\langle p^{\lambda}p^{\rho}  p^{\nu}\right\rangle\beta_{\nu,\mu}+ \Delta^{\mu}_{\rho} \left\langle p^{\lambda}p^{\rho} \chi_{n,\mu}\right\rangle
\label{che4b}
\te
Solving these equations (see Appendixes (\ref{gen-1pdf-0order}) and (\ref{gen-1pdf-2order})) we find the first order

\begin{equation}
 \chi_1 =  - \frac{p^{\mu}p^{\rho}}{F\left[-\beta_{\nu}p^{\nu}\right]}
 \frac{\sigma_{\mu\rho}}{2t^3} \label{chi-1-2t}
\end{equation}
where $\sigma_{\mu\rho}$ is the shear tensor
\begin{equation}
 \sigma_{\mu\nu} = \Lambda_{\mu\nu}^{\alpha\beta} u_{\alpha,\beta}\label{shear}
\end{equation}

\begin{equation}
 \Lambda_{\mu\nu}^{\alpha\beta} = \left[\Delta_{\mu}^{\alpha}\Delta_{\nu}^{\beta}
 + \Delta_{\nu}^{\alpha}\Delta_{\mu}^{\beta}
 - \frac{2}{3}\Delta_{\mu\nu}\Delta^{\alpha\beta}\right]\label{projector-1}
\end{equation}

{If we were to stop at this order and take $F = const.$ we would parameterize $\chi$ with just the function $X_1=\Lambda^{\mu\nu}_{\rho\sigma}p^{\rho}p^{\sigma}/const$.} This would lead us to Israel-Stewart theory, which is not satisfactory; in particular, it yields no dynamics for tensor modes \cite{PeCa21}. {This is one of the reasons why} we shall go further one more order

\bea 
\chi_2&=&\frac1{t^2}\Sigma_{\nu\mu\rho}\frac{p^{\nu}p^{\mu}p^{\rho}}{2t^3F^2}-\frac1{t^3}u_{\mu}t_{,\lambda\tau}\Lambda_{\rho\nu}^{\lambda\tau}\frac{p^{\nu}p^{\mu}p^{\rho}}{2t^3F^2}\nn
&+&\frac1{5t^3}\sigma^{\rho}_{\nu,\rho}p^{\nu}\left[\frac {D_2}F+\frac{\Theta^1_4}{\Theta^0_3}\frac{p^{\mu}u_{\mu} }{tF}+\left(\frac{p^{\mu}u_{\mu} }{tF}\right)^2-D_3\right] \label{chidos}
\tea
The tensor $\Sigma_{\sigma\rho\lambda}$ is found from the decomposition

\begin{eqnarray}
 \Delta^{\lambda}_{\tau} p^{\tau}p^{\sigma}p^{\rho} \sigma_{\sigma\rho,\lambda} &=& 
   p^{\lambda}p^{\sigma}p^{\rho}\Sigma_{\sigma\rho\lambda}
  + \frac{2}{5}\left(p^{\mu}u_{\mu}\right)^2 p^{\sigma}\sigma^{\rho}_{\sigma,\rho}
  \label{derivada-sigma}
 \end{eqnarray}
Where $\Sigma_{\sigma\rho\lambda}$ is transverse, symmetric and traceless, 

\be
D_2=\frac{\Theta^1_4\Theta^1_3}{\Theta^0_3\Theta^1_2}-\frac{\Theta^2_4}{\Theta^1_2}
\label{dedos}
\te
and

\begin{equation}
 D_3 = \left[D_2\frac{\Theta^1_3}{\Theta^0_3}-\left(\frac{\Theta^1_4}{\Theta^0_3}\right)^2
 +\frac{\Theta^2_5}{\Theta^0_3}\right]
 \label{detres}
\end{equation}
with the functions

\be 
 \Theta_{n}^{m} = \frac{1}{2\pi^2}\int du ~ \frac{u^{n+1}}{F^m[u]} e^{-u}\label{Thetanm}
\te

We observe that the whole second line of equation (\ref{chidos}) vanishes when the Anderson Witting collision term is chosen. This was to be expected because in this limit these terms collapse into two terms proportional to $p^{\mu}$ and to $p^{\mu}/u_{\rho}p^{\rho}$ which do not show up when the Anderson-Witting prescription $F[x]=x$ is used from scratch.

This concludes the construction of the second order Ch-En solution. We now must use $\chi_1$ and $\chi_2$ as templates whereby to identify which functions of momentum to include into a general parameterization of $\chi$.

 \subsection{Dynamics from the moments approach}\label{grad}

 Assuming a parameterization of the form eq. (\ref{chihydro}) begs the question of which functions $X^{\alpha}$ ought to be included. In search of guidance, we look at the functions of momentum which actually show up in the second order Ch-En solution. From eqs. (\ref{chi-1-2t}) and (\ref{chidos}) we see that the second order $\chi$ may be regarded as a linear combination of four tensor fields
 
 \bea
X^1&=& \frac1{2t^2F}\Lambda^{\mu\rho}_{\lambda\tau}p^{\lambda}p^{\tau}\nn
X^2&=&\frac1{6t^3F^2}\Lambda^{\mu\nu\rho}_{\lambda\sigma\tau}p^{\lambda}p^{\sigma}p^{\tau}\nn
X^3&=& \frac1{2t^2F}\Lambda^{\mu\rho}_{\lambda\tau}p^{\lambda}p^{\tau}\left[-\frac{u_{\mu}p^{\mu}}{tF}-\gamma\right]\nn
X^4&=&\frac1t\Delta^{\nu}_{\mu}p^{\mu}\left[\frac{D_2}{F}+\frac{\Theta^1_4}{\Theta^0_3}\frac{p^{\mu}u_{\mu} }{tF}+\left(\frac{p^{\mu}u_{\mu} }{tF}\right)^2-D_3\right] \label{xlist}
\tea
where 

\be
\gamma=\frac{\Theta^3_6}{\Theta^2_5}
\label{gamma}
\te
and $\Lambda^{\mu\nu\rho}_{\lambda\sigma\tau}$ is the projection over transverse, totally symmetric and traceless third order tensors. The coefficients of the linear combination 
are the four tensors $C_{\alpha}$ that represent the different parameters of
the theory and have the same symmetry properties as the corresponding coefficients in eqs. (\ref{chi-1-2t}) and (\ref{chidos}). 
The $X^{\alpha}$ tensors are totally symmetric, traceless and transverse with respect to the Landau-Lifshitz velocity, and moreover they obey the orthogonality condition

\be 
\left\langle p^0X^{\alpha}X^{\beta}\right\rangle\propto\delta^{\alpha\beta}
\label{ortho-chi}
\te
  The equations of motion are the moment equations (\ref{moments})
where $\chi=\sum_{\alpha=1}^4C_{\alpha}X^{\alpha}$, which  is consistent with the constraint eq. (\ref{constraint}), and $f_0=e^{\beta_{\mu}p^{\mu}}$. The equation corresponding to $\alpha=0$ is just energy-momentum conservation.

\subsection{The complete set of equations of motion}

As our ultimate goal is to compute propagation speeds, we develop here the linear form of the transport part of the equations  (\ref{moments}) for the different $C_{\alpha}$'s. We begin by writing

\be
f_{,\mu}=f_0\left(\beta_{\nu,\mu}p^{\nu}+\sum_{\alpha=1}^4C_{\alpha,\mu}X^{\alpha}\right)
\label{1pdf-deriv}
\te 

The moments of the transport term include energy-momentum conservation

\be
\left\langle p^{\lambda}p^{\mu}\left(\beta_{\nu,\mu}p^{\nu}+\sum_{\alpha=1}^4C_{\alpha,\mu}X^{\alpha}\right)\right\rangle=0 \label{gen-EMT-eq-fs}
\te 
and the moment equations 

\be
\left\langle X^{\gamma}p^{\mu}\left(\beta_{\nu,\mu}p^{\nu}+\sum_{\alpha=1}^4C_{\alpha,\mu}X^{\alpha}\right)\right\rangle=\ldots \label{gen-mom-fs}
\te
with $\gamma=1, \ldots, 4$. {The right hand sides of eqs. (\ref{gen-mom-fs}) are immaterial because they contain no derivative terms, while propagation speeds are defined by the principal terms in the equations \cite{HL83} }. 

Working in the rest frame where $u^{\mu}=(1,0,0,0)$ and setting $t=1$ we have

\bea
\left\langle p^{0}p^{\mu}p^{\nu}\right\rangle\beta_{\nu,\mu}+\sum_{\alpha=1}^4\left[\dot C_{\alpha}\left\langle \left(p^{0}\right)^2 X^{\alpha}\right\rangle+C_{\alpha,i}\left\langle p^{i}p^{0}X^{\alpha}\right\rangle\right]&=&0 \nn
\left\langle p^{j}p^{\mu}p^{\nu}\right\rangle\beta_{\nu,\mu}+\sum_{\alpha=1}^4\left[\dot C_{\alpha}\left\langle p^{0}p^jX^{\alpha}\right\rangle+C_{\alpha,i}\left\langle p^{i}p^{j}X^{\alpha}\right\rangle\right]=0 \nn
\left\langle X^{\gamma}p^{\mu}p^{\nu}\right\rangle\beta_{\nu,\mu}+\sum_{\alpha=1}^4\left[\dot C_{\alpha}\left\langle X^{\gamma}p^{0} X^{\alpha}\right\rangle+C_{\alpha,i}\left\langle X^{\gamma}p^{i} X^{\alpha}\right\rangle\right]&=&0 \label{1st-set}
\tea

The constraint eq. (\ref{constraint}) implies that $\left\langle (p^{0})^2 X^{\alpha}\right\rangle=\left\langle p^{0}p^jX^{\alpha}\right\rangle=0$, so equations 
(\ref{1st-set}) simplify to

\begin{eqnarray}
\left\langle p^{0}p^{\mu}p^{\nu}\right\rangle\beta_{\nu,\mu}&=&0\label{cons-eq-1}\\
\left\langle p^{j}p^{\mu}p^{\nu}\right\rangle\beta_{\nu,\mu}+\sum_{\alpha=1}^4C_{\alpha,i}\left\langle p^{i}p^{j}X^{\alpha}\right\rangle&=&0 \label{cons-eq-2}\\
\left\langle X^{\beta}p^{i}p^{j}\right\rangle\beta_{i,j}+\sum_{\alpha=1}^4\left[\dot C_{\alpha}\left\langle X^{\beta}p^{0} X^{\alpha}\right\rangle+C_{\alpha,i}\left\langle X^{\beta}p^{i} X^{\alpha}\right\rangle\right]&=&0
\label{cons-eq-3}
\end{eqnarray}
Computing the averages as defined in eq. (\ref{MV-def})
we find the full set of equations in the free-streaming regime 

\begin{eqnarray}
 \frac{\dot t}{t} + \frac{1}{3} u^j_{,j}  &=& 0 \nn
 \Theta_3^0 \left( \frac{t_{,i}}{t}+\dot u^i \right)
 + \frac{1}{5} \Theta_4^1\Lambda^{\mu\rho ij} C_{1\mu\rho,j}
 + \frac{1}{5} A \Lambda^{\mu\rho i j} C_{3\mu\rho,j} &=& 0 \nn
 \Theta_4^1 \Lambda^{\mu\nu kl} u_{k,l} + \Theta_5^2 \Lambda^{\mu\nu kl} \dot C_{1 kl}
 + \frac{1}{7} \Theta_6^3 \Lambda^{\mu\nu k lmn } C_{2 lmn ,k}
 +  F \Lambda^{\mu\nu\lambda k} C_{4\lambda ,k} &=& 0\nn
 \Theta_6^3 \Lambda^{\mu\nu\lambda\rho\sigma k} C_{1\rho\sigma,k}+\Theta_7^4 \Lambda^{\mu\nu\lambda\rho\sigma\tau} \dot C_{2\rho\sigma\tau} 
  +  G \Lambda^{\mu\nu\lambda\rho\sigma k} C_{3\rho\sigma,k} &=& 0 \nn
  A \Lambda^{\mu\nu kl} u_{k,l} + \frac{1}{7} G \Lambda^{\mu\nu klmn } C_{2 lmn,k}
  +  G \Lambda^{\mu\nu kl} \dot C_{3kl} 
  +   K \Lambda^{\mu\nu l k} C_{4 l,k} &=& 0 \nn
 \frac{1}{5} F \Lambda^{\mu\nu \rho k} C_{1 \nu \rho,k}
 + \frac{1}{5} K \Lambda^{\mu\nu \rho k} C_{3 \nu \rho,k}+ M \Delta^{\mu\nu} \dot C_{4 \nu}   &=& 0 
 \label{fruttitutti}
\end{eqnarray}

The different coefficients in the equations read

\begin{eqnarray}
 A &=& \left[\Theta_5^2 - \gamma \Theta_4^1\right]\nn
 F &=& \left[D_2\Theta_4^2 -\frac{\Theta^1_4}{\Theta^0_3} \Theta_5^2 + \Theta_6^3 - D_3 \Theta_4^1\right]\nn
 G &=&  \left[\Theta_7^4 - \gamma \Theta_6^3 \right]\nn
 K &=& \left[D_2\Theta_5^3 - \frac{\Theta^1_4}{\Theta^0_3}\Theta_6^3 + \Theta_7^4 - D_3 \Theta_5^2 
 -\gamma F\right]\nn
 M &=& \left[D_2^2 \Theta_3^2 + \left(\frac{\Theta^1_4}{\Theta^0_3}\right)^2 \Theta_5^2 + \Theta_7^4
 + D_3^2 \Theta_3^0 - 2 D_2 \frac{\Theta_4^1}{\Theta_3^0} \Theta_4^2 
 - 2 \frac{\Theta_4^1}{\Theta_3^0} \Theta_6^3 + 2 D_2 \Theta_5^3
 \right.\nn
 &-& \left. 2 D_2 D_3 \Theta_3^1 + 2 D_3 \frac{\Theta_4^1}{\Theta_3^0} \Theta_4^1 - 2 D_3 \Theta_5^2  \right] \label{coef-frutti}
\end{eqnarray}

Observe that these coefficients depend explicitly on the microphysics, due to the
presence of the functions (\ref{Thetanm}). In particular the coefficients 
$D_2$, $D_3$ and $\gamma$, defined in eqs. (\ref{dedos}), (\ref{detres}) and (\ref{gamma}), respectively, vanish in the AW ansatz for the RTA.

\section{Propagation speeds}\label{ps}

We are now ready to derive the propagation speeds $v$ for linearized fluctuations around an equilibrium solution $\beta_{\mu}=\beta_{0\mu}=$ constant, $C^{\alpha}=0$, $\alpha=1-4$. These fluctuations represent collective modes of the system of interacting particles. {As we shall show presently, the propagation speeds $v$ may be derived from the dispersion relations obtained from the
Fourier transform of 
equations (\ref{fruttitutti}), which take the form $\omega=vk$. To derive this dispersion relations from the principal terms only is equivalent to consider the full dispersion relations, including dissipative terms, in the limit $k\to\infty$.}

It is important to highlight that {working with the equation system (\ref{fruttitutti})}
does not
make the choice of the function $F$ in the collision term irrelevant, because the tensors $X^{\alpha}$ which make up $\chi$ depend on it, see eqs. (\ref{xlist}) and (\ref{coef-frutti}). This has stemmed from enforcing the integrability conditions at each order in the Ch-En development, which are essential for the second law of thermodynamics to be fulfilled. 
So we have written a parameterized 1-pdf general enough to include a second order Ch-En solution of the collisional theory, and {work with equations (\ref{fruttitutti})} to investigate the front propagation speeds. We shall find that the propagation speeds depend on the microphysics through the choice of the function $F$, and for a wide range of choices, are maximized by the AW ansatz $F[x]=x$.

\subsection{Relating front propagation speeds to the {dispersion relations}}

Regardless of the actual form of the collision term, it is clear that after linearizing around an equilibrium solution and going to the equilibrium rest frame, the equations (\ref{moments}) will take the form

\be 
\dot C^{\alpha}+N^{j\alpha}_{\beta}C^{\beta}_{,j}+I^{\alpha}_{\beta}C^{\beta}=0
\label{fulleoms}
\te 
where only the last term comes from the collision integral.

Let us seek a solution representing a front (namely a surface where the fluid variables are continuous but their first derivatives are not) moving into fluid in equilibrium along the direction $\hat{\mathbf{k}}$. The solution depends on time and space only through the variable $\xi=\hat{\mathbf{k}}\cdot\mathbf{x}-vt$, where $v$ is the front velocity. Let the front position be $\xi=\xi_0$. At this point, the $C^{\alpha}$'s are continuous but the lateral $\xi-$ derivatives ${C'}^{\alpha}_+$ and ${C'}^{\alpha}_-$ are different ( $+$ and $-$ denote the upstream and downstream parts of the fluid). Therefore, taking the difference of the eqs. (\ref{fulleoms}) in front of and behind the front, the terms from the collision integral cancel and we get

\be 
\left[-v\delta^{\alpha}_{\beta}+\hat{\mathbf{k}}_jN^{j\alpha}_{\beta}\right]({C'}^{\beta}_+-{C'}^{\beta}_-)=0
\label{propspeed}
\te  
with a prime denoting a $\xi$- derivative. 

On the other hand suppose we seek the dispersion relations which follow from eqs. (\ref{fulleoms}). Then we propose a solution of the form $C^{\alpha}=C^{\alpha}_0e^{i\left[\mathbf{k}\cdot\mathbf{x}-\omega t\right]}$ and obtain

\be 
[(-i\omega ) \delta^{\alpha}_{\beta}+iN^{j\alpha}_{\beta}\mathbf{k}_{j}+I^{\alpha}_{\beta}]C^{\beta}_0=0
\label{disprel}
\te 
It is clear that eqs. (\ref{propspeed}) are the same as eqs. (\ref{disprel}) under the identification $\omega=vk$, $\mathbf{k}=k\hat{\mathbf{k}}$. We shall take advantage of this {fact and evaluate the propagation speeds from the Fourier
transform of system (\ref{fruttitutti})}. This is not an approximation but rather the definition of the propagation speeds.

\subsection{SVT Decomposition}\label{svt}

As it is well known, the equations of motion can be further decoupled by decomposing the deviations from equilibrium into scalar, vector and tensor quantities. We then write

\begin{eqnarray}
 u^i &=& \vartheta^i + \nabla_i \vartheta \label{veli-dec} \\
 C_{4i} &=& c_{4i} + \nabla_i c_4 \label{C4i-dec}\\
 C_{1ij} &=& \left( \nabla_i \nabla_j - \frac{1}{3}\Delta_{ij}\nabla^2\right) c_1 + c_{1i,j} + c_{1j,i} + c_{1ij} \label{c1ij-dec}\\
 C_{3ij} &=& \left( \nabla_i \nabla_j - \frac{1}{3}\Delta_{ij}\nabla^2\right) c_3 + c_{3i,j} + c_{3j,i} + c_{3ij} \label{c3ij-dec}\\
 C_{2ijk} &=& \left[\nabla_i \nabla_j \nabla_k - \frac{1}{5}\left(\Delta_{ij}\nabla_k + \Delta_{ik}\nabla_j
  + \Delta_{jk}\nabla_i\right) \nabla^2\right] c_2 \nn
  &+& \left( \nabla_i \nabla_j - \frac{1}{5}\Delta_{ij}\nabla^2\right)c_{2k}
  + \left( \nabla_i \nabla_k - \frac{1}{5}\Delta_{ik}\nabla^2\right)c_{2j}
  + \left( \nabla_k \nabla_j - \frac{1}{5}\Delta_{kj}\nabla^2\right)c_{2i} \nn
  &+& \nabla_i c_{2jk} + \nabla_j c_{3ik} + \nabla_k c_{2ij} + c_{2ijk} \label{c2ijk-dec}
\end{eqnarray}
Here quantities with no indexes are scalars, with one single index are vectors (namely divergenceless) and with more that one tensors (divergenceless and traceless). 

We shall discuss in some detail the simplest tensor sector, and give the results for the vector and scalar ones, which follow the same structure. {The homogeneous} equation for $ c_{2ijk}$ reduces to $\dot c_{2ijk}=0$ and we shall not discuss it.

\subsubsection{Tensor Sector}\label{Tsector}

Considering only tensor quantities in eqs. (\ref{fruttitutti}) we get

\begin{eqnarray}
 \Theta_5^2 \dot C_{1 ij} + \frac{3}{7} \Theta_6^3 \nabla^2 C_{2ij} &=& 0\nn
 \Theta_6^3  C_{1ij} +3 \Theta_7^4 \dot C_{2ij} +  G C_{3ij} &=& 0 \nn
  G [\frac{3}{7} \nabla^2 C_{2ij}+\dot C_{3ij} ] &=& 0 \label{tens-eq}
\end{eqnarray}
Therefore the dispersion relation is derived from the roots of the determinant

\be 
\mathrm{det}\;\left(\begin{array}{ccc}
\Theta_5^2 (-i\omega) & -\frac{3}{7} \Theta_6^3 k^2  & 0\\
 \Theta_6^3  &3 \Theta_7^4 (-i\omega) &  G  \\
 0&  -\frac{3}{7}G k^2 & G(-i\omega)\end{array}\right)=0
 \label{tensorfrutti}
 \te
Explicitly

\be
3G\Theta^2_5\Theta^4_7(i\omega)[\omega^2-\frac17k^2]=0 \label{disp-rel-tens}
\te 
Observe that although the determinant in eq. (\ref{tensorfrutti}) becomes singular for the AW RTA, where $G=0$, the dispersion relation is well defined there, and yields the propagation speeds $1/\sqrt{7}$ and $0$, independently of the choice of the function $F$, in agreement with \cite{PeCa21}.

As we shall show presently, the propagation speeds do depend on the choice of the function $F$ in the vector and scalar sectors.

\subsubsection{Vector Sector}\label{Vsector}

The vector terms in eq. (\ref{fruttitutti}) are

\begin{eqnarray}
 \Theta_3^0 \dot \vartheta^i + \frac{2}{5} \Theta_4^1 \nabla^2 c_{1i} + \frac{2}{5} A \nabla^2 c_{3 i} &=& 0 \nn
 \nabla_j\left[\Theta_4^1 \vartheta_{i} + 2\Theta_5^2 \dot c_{1 i}
 + \frac{24}{35}\Theta_6^3 \nabla^2 c_{2i} +  F  c_{4 i}\right] &=& 0\nn
  \left(\nabla_j\nabla_k - \frac{1}{5}\Delta_{jk}\nabla^2\right)\left[\frac{2}{3}\Theta_6^3  c_{1i}  + \Theta_7^4  \dot c_{2 i} 
  + \frac{2}{3} G c_{3i}\right] &=& 0 \nn
 \nabla_j \left[A \vartheta_{i}+ \frac{24}{35} G \nabla^2 c_{2i} + 2 G \dot c_{3i} 
  + K  c_{4i} \right] &=& 0 \nn
  \frac{2}{5} F \nabla^2 c_{1i} + \frac{2}{5} K \nabla^2 c_{3 i} +M \dot c_{4 i}  &=& 0 
  \label{vec-eq-set}
\end{eqnarray}
wherefrom we get the characteristic equation

\be 
\mathrm{det}\;\left(\begin{array}{ccccc}
\Theta_3^0 (-i\omega) & -\frac{2}{5} k^2\Theta_4^1&0 &-\frac{2}{5}k^2 A&0 \\
 \Theta_4^1 & 2\Theta_5^2 (-i\omega) & -\frac{24}{35}k^2\Theta_6^3  &0&  F  \\
  0&\frac{2}{3}\Theta_6^3  & \Theta_7^4  (-i\omega) &\frac{2}{3} G &0\\
 A &0&- \frac{24}{35} G k^2&2 G (-i\omega) & K  \\
  0&-\frac{2}{5} F k^2&0&- \frac{2}{5} K k^2&M (-i\omega) 
\end{array}\right)=0
 \label{vectorfrutti}
 \te

\subsubsection{Scalar Sector}\label{Ssector}

Keeping only scalar terms in eq. (\ref{fruttitutti}) gives
\begin{eqnarray}
 \dot t + \frac{1}{3} \nabla^2 \vartheta  &=& 0 \nn
 \nabla_i\left[\Theta_3^0 \left(t+\dot \vartheta \right)
 + \frac{4}{15} \Theta_4^1 \nabla^2 c_1 + \frac{4}{15} A \nabla^2 c_3\right] &=& 0 \nn
 2\left(\nabla_i\nabla_j - \frac{1}{3}\Delta_{ij}\nabla^2\right)\left[\Theta_4^1 \vartheta
 +  \Theta_5^2  \dot c_{1}
 +  \frac{9}{35}\Theta_6^3 \nabla^2c_2
 +  F c_4 \right] &=& 0\nn
 6 \left[ \nabla_i \nabla_j \nabla_k - \frac{1}{5} \nabla^2 \Delta_{ij}\nabla_k
 - \frac{1}{5} \nabla^2 \Delta_{ik}\nabla_j - \frac{1}{5} \nabla^2 \Delta_{jk}\nabla_i \right]
 \left[\Theta_6^3 c_1 + \Theta_7^4 \dot c_{2} +  G c_3\right] &=& 0 \nn
  2\left(\nabla_i\nabla_j - \frac{1}{3}\Delta_{ij}\nabla^2\right)\left[ A \vartheta + \frac{9}{35}  G  \nabla^2c_2
  +  G \dot c_{3} +  K c_{4}\right] &=& 0 \nn
 \nabla_i \left[  \frac{4}{15} F \nabla^2 c_{1}
 + \frac{4}{15} K \nabla^2 c_{3}+M \dot c_{4} \right] &=& 0 \label{scal-eq-set}
\end{eqnarray}
leading to the dispersion relations

\be
\mathrm{det}\;\left(\begin{array}{cccccc}
(-i\omega)  & -\frac{1}{3} k^2  &0  &0 &0 &0\\
\Theta_3^0  &\Theta_3^0 (-i\omega)  
 & -\frac{4}{15} \Theta_4^1 k^2  &0 & -\frac{4}{15} A k^2 &0\\
 0&\Theta_4^1 
 &  \Theta_5^2  (-i\omega)
 &  -\frac{9}{35}\Theta_6^3 k^2 &0
 &  F  \\
0 &0&\Theta_6^3  & \Theta_7^4 (-i\omega) &  G   &0\\
0& A&0 & -\frac{9}{35}  G  k^2
  &  G (-i\omega)  &  K \\
 0& 0 &-\frac{4}{15} F k^2  &0
 & -\frac{4}{15} K k^2 &M (-i\omega) 
\end{array}\right)=0
\label{scalarfrutti}
\te

\subsection{Results}

To give some content to the results above, we shall consider the family of RTA's where
\be 
F[x]=x^a \label{F-ansatz}
\te
for which
\be 
\Theta^n_m=\Gamma[2+m-an] \label{Theta-ansatz}
\te 
To avoid infrared divergences in the equations of motion we require $a\le 2$. This family includes the Marle and AW RTAs as particular cases, namely $a=0$ and $a=1$, respectively.

As in the tensor case, the dispersion relations are well defined at $a=1$, although the matrices in eqs. (\ref{vectorfrutti}) and (\ref{scalarfrutti}) are singular there. The propagation speeds for $a=1$ coincide with the values given in \cite{PeCa21}.

For the vector and scalar sectors we used the tool "Mathematica" to solve the dispersion relations (\ref{vectorfrutti}) and (\ref{scalarfrutti}) and plotted the solutions in the figures below.  

\begin{figure}[ht]
\centering
\includegraphics[scale=0.8]{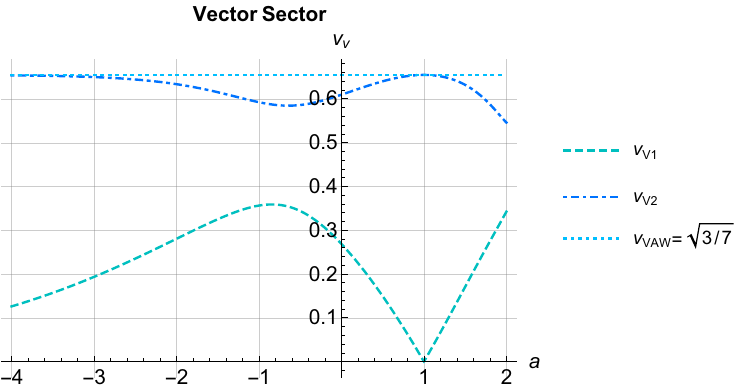}
\caption{(Color online) Speeds of the two vector modes from eq. (\ref{vectorfrutti}). The fastest mode (top dot-dashed blue curve) attains the same maximum speed for $a=1$ and for $a\to -\infty$ (top dotted light-blue horizontal line), indicating that the AW value is not exceeded at any value of $a$. Observe that the slowest mode speed (bottom, dashed light blue curve) is zero for $a=1$, so we recover the AW case \cite{PeCa21} where only one non-null mode exists.}
\label{vs}
\end{figure}

The solutions $v_V = \omega/k$ for the vector sector are plotted in Fig. (\ref{vs}). 
The fastest mode (top dot-dashed blue curve) attains the same maximum speed for $a=1$ and for $a\to -\infty$ (top dotted light-blue horizontal line), indicating that the AW value is not exceeded at any value of $a$. Observe that the slowest mode speed (bottom, dashed light blue curve) is zero for $a=1$, so we recover the AW case where only one non-null mode exists \cite{PeCa21}.

\begin{center}
\begin{figure}[ht]
\includegraphics[scale=0.8]{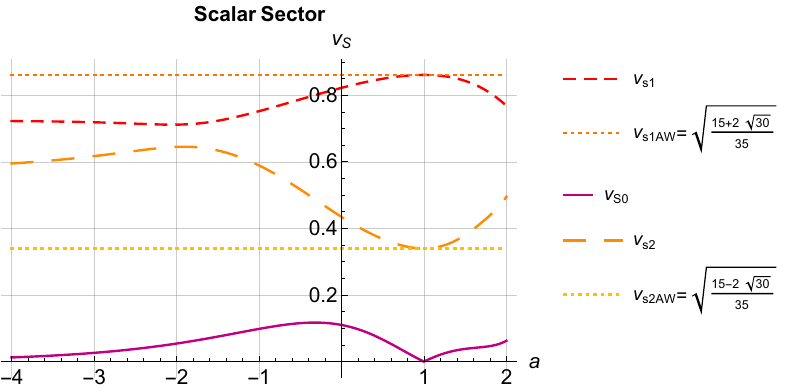}
\caption{(Color online) Speeds of the three scalar modes from eq. (\ref{scalarfrutti}). 
We see that the maximum speed of the fastest mode (top red short-dashed curve) corresponds to the AW solution  $a=1$ (top horizontal orange dotted line). The intermediate speed mode
(long-dashed orange curve in the middle of the figure) attains its minimum value also at the AW value $a=1$ (bottom horizontal yellow dotted line), and the speeds of this mode never exceed the ones of the fastest mode. These two modes are the generalization of the 
AW modes found elsewhere. The bottom, single-line purple curve corresponds to the speeds of a new, slowest mode, whose velocity for $a=1$ is zero. Thus we see that the AW case, for which there are only two propagating modes, is consistently included in our formalism.}
\label{ss}
\end{figure}
\end{center}

The solutions $v_S = \omega/k$ for the scalar case are plotted in Fig. (\ref{ss}). We see that the maximum speed of the fastest mode (top red short-dashed curve) corresponds to the AW solution  $a=1$ (top horizontal orange dotted line). The intermediate speed mode
(long-dashed orange curve in the middle of the figure) attains its minimum value also at the AW value $a=1$ (bottom horizontal yellow dotted line), and the speed of this mode never exceeds the one of the fastest mode. These two modes are the generalization of the 
AW modes found elsewhere. The bottom, single-line purple curve corresponds to the speeds of a new, slowest mode, whose velocity for $a=1$ is zero. Thus we see that the AW case \cite{PeCa21}, for which there are only two non-null modes, is consistently included in our formalism.

\begin{center}
\begin{figure}[ht]
\includegraphics[scale=0.8]{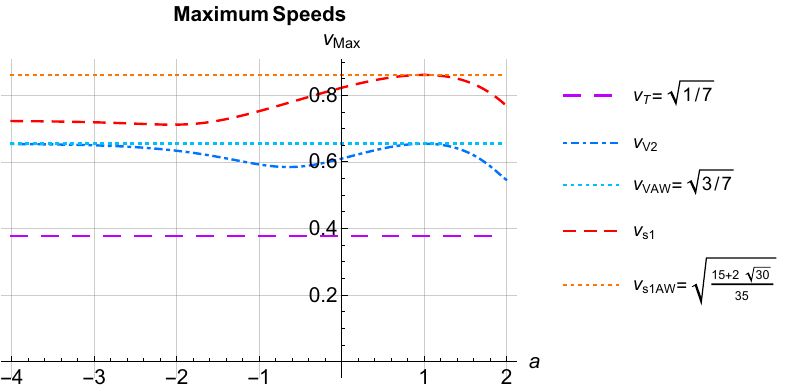}
\caption{(Color online) Comparison of the maximum propagation speeds of each sector. Top dotted horizontal line corresponds to the AW ($a=1$) scalar mode speed, short-dashed line immediately below corresponds to the velocities of the scalar fastest mode of our model. Middle dotted horizontal line and dot-dashed middle curve correspond respectively to the vector mode speed for AW ($a=1$) and to the speeds of our model. Bottom long-dashed horizontal line is the speed of the tensor mode, which agrees with the AW speed over the entire interval of $a$ values considered. They verify $v_T < v_V < v_S$.}
\label{ms}
\end{figure}
\end{center}

Finally, in Fig. (\ref{ms}) we plotted the curves that correspond to the speed of the 
fastest mode of each sector. The
top dotted horizontal line corresponds to the AW ($a=1$) speed, short-dashed line immediately below corresponds to the velocities of the scalar fastest mode of our model. Middle dotted horizontal line and dot-dashed middle curve correspond respectively to the vector mode speed for AW ($a=1$) and to the speeds of our model. Bottom long-dashed horizontal line is the speed of the tensor mode, which agrees with the AW speed over the entire interval of $a$ values considered. They verify $v_T < v_V < v_S$,
{which curiously is the same order relationship already obtained by Israel and Stewart in
Ref. \cite{IS79a}.}

{All the propagation speeds are below the speed of the light. This is a consequence of the causal evolution of the Boltzmann equation, which is something that is broken when performing the Chapman-Enskog 
procedure (relativistic Navier-Stokes is acausal and unstable) and is recovered by (1-pdf) moment methods such as the one of the present paper. For further discussion of causality requirements in relativistic hydrodynamics see Ref. \cite{GDN24} .}

\section{Conclusions}\label{conc}

Using the Chapman-Enskog expansion, we developed a linearized 1-pdf up to second order around local thermal equilibrium. At each order we enforced the second law, so positive entropy production is guaranteed. We then generalized this distribution function by identifying each term in the Ch-En expansion with a product of a momentum-dependent tensor with a parameter, that encodes the dissipative properties of the flow. In this way the parameterization contains the second order Ch-En solution as a special case. Using the moments method we then obtained the linearized equations for scalar, vector and tensor perturbations. We worked with the RTA for the collision integral and considered the relaxation time
as a function of the momentum, $F\left[-\beta_{\mu}p^{\mu}\right]$. 

The coefficients of the conservation equations depend on the function $F$. 
They form a family of parameterized theories that describe different phenomenologies 
depending on the choice of the function $F$. Thus {in equations (\ref{fruttitutti})} there remains information about the microphysics on which the RTA was built.
Stated in other words: The choice of $F$ is a crucial part of the construction of the RTA.

To analyze a concrete case, we specialized the general equations to the case where $F=\left(\beta_{\mu}p^{\mu}\right)^{a}$, which includes the choices of AW $a=1$ \cite{AWa,AWb} and Marle $a=0$ \cite{Marle-a,Marle-b}, widely used in the literature, as particular cases. 
{The choice of a power-law, besides being mathematically tractable, has actually been
proposed before in the context of relativistic heavy ion collisions from phenomenological considerations
\cite{DMT10,LO10,RDN21,DeNo24,RFDN22,PRCal10,PRCal13,KW19,WW21,M21,M22,AM23,AM24,H23,DBJJ22,DBJJ23,B23}.
For example, the interpolating values $0\le a\le 1$ have already been discussed in Refs. \cite{DMT10,LO10}, motivated by its possible application to improve the description of relativistic heavy ion collisions. 
Here we have included the full range $-\infty\le a\le 2$, which is maximal because larger values of $a$ leads to infrared divergences in the coefficients of the hydrodynamic equations. Allowing for a negative $a$ allows us to explore distributions heavily biased towards hard modes. 
The power of the generalized RTA presented here is the ability to reproduce spectral properties of the kinetic equation, most importantly whether zero is an isolated eigenvalue or it is embedded in the continuous spectrum.   Figs. (\ref{vs}) and (\ref{ss}) show the dependence of the propagation speeds on the choice of $F$, restricted to a power law. Choosing a more general functional form for $F$ may be justified by concrete experimental results and/or deduced from the Boltzmann equation
\cite{LY16,DMT10,DeNo24,DBJJ23,RBD23b}.}

The propagation speeds of a theory are fundamental to determine causality and to discuss shock waves, among other effects. 
The linear conservation equations decouple into three sets, corresponding to the tensor, vector and scalar modes and we computed the corresponding propagation speeds. The propagation speeds are the phase velocity for plane waves {obtained from Eqs. (\ref{fruttitutti})}, and as it was emphasized above, they depends on the choice of $F$.

For the given 1-pdf the number of tensor, vector and scalar modes are respectively 2, 5 
and 6. 

For the tensor modes we found the propagation speed is actually independent of $a$ and agrees with the AW value \cite{PeCa21}. 

In the vector sector, besides the trivial solution $v_v=0$, there are two propagation speeds shown in Fig. (\ref{vs}). There we saw that the fastest propagation speed is bounded above by the AW value \cite{PeCa21}, which is reached at $a=1$. The slower mode has $v_v=0$ for $a=1$. Therefore the number of dynamical vector modes in the AW limit reduces to $2$, as it must.

For the scalar sector we obtained three different propagation speeds as is shown in Fig. 
(\ref{ss}). As in the tensor and vector sectors, the fastest mode has maximum velocity at 
$a=1$, where we recover the AW result \cite{PeCa21}. For the intermediate mode we also recover the lower AW speed when $a=1$. The speeds of the 
slowest mode are significantly lower than the other two scalar modes, and vanishes for $a=1$. In this way we recover the right number of dynamical scalar modes ($4$) in the AW case.

In Fig. (\ref{ms}) we compared the speeds of the three fastest modes. We see that
 they satisfy $v_T < v_v < v_s$ throughout the whole range of $a$ values \cite{BR99}. 
 {In Ref. \cite{IS79a}  Israel and  Stewart also calculated the propagation speeds
 for scalar, vector and tensor modes and found the same order relationship obtained in this work.}

 {We expect that including higher orders in the Chapman-Enskog development, besides adding more functions to the 1-pdf parameterization, will produce increasingly higher speeds, which will asymptotically approach the speed of light, as was demonstrated by G. Boillat, T. Ruggieri and I. M\"uller \cite{BR99,BR79,BR97a,BR97b,Mull99}.}

 {As stated in Subsection \ref{gen-RTA-appr} in this work we have worked in the Landau frame throughout. The frame-dependence of hydrodynamics, and the ensuing possibility of improving the hydrodynamic description by a judicious choice of frame, are active areas of research 
 \cite{DoGaMoShTo22,BhaMiRoSi24,BhaMiRo24} We intend to provide a deeper analysis of the frame dependence of the results in this paper in forthcoming work.}

We believe that the main contributions of this work are, first, the use
of Chapman-Enskog expansion as a template on which to build a parameterized theory with a dynamics based on the
method of moments. The resulting theory is causal for the full range of values of $a$. Causality is expected because, as we have already said, the theory is built to enforce thermodynamic stability, and it is known that stability, causality and covariance are closely linked \cite{SM22,HSSW23,ASN23,BRD23,HK24,WP24,GDN24}. 
Second, that the fastest propagation speeds are found in the AW limit $a=1$, for all scalar, vector and tensor modes. To the best of our knowledge, the fact that the Anderson-Witting RTA \cite{AWa,AWb} produces the fastest propagation speeds is a new result. This has deep implications for the description of strong shocks in relativistic fluids \cite{Cal22}, which we expect to elaborate on in a separate contribution.

\section*{Acknowledgements}

E. C. acknowledges financial support from Universidad de Buenos Aires through Grant No. UBACYT
20020170100129BA, CONICET Grant No. PIP2017/19:11220170100817CO and ANPCyT Grant No. PICT 2018: 03684. A.K. acknowledges financial support through project UESC 073.11157.2022.0001594-04.

\appendix

\section{Chapman-Enskog solution}\label{Chensol}

\subsection{First Order Ch-En solution}\label{gen-1pdf-0order}

We obtain the first order Ch-E solution setting $n=0$ in eqs. (\ref{che3b}) and (\ref{che4b}), whereby

\be
{Q}\left[t^2 F\left[-\beta_{\nu}p^{\nu}\right]  {Q}\left[\chi_{1}\right]\right]=p^{\mu}u_{\mu} p^{\nu}\dot\beta_{\nu}^{\left(0\right)}-p^{\rho} \Delta^{\mu}_{\rho} p^{\nu}\beta_{\nu,\mu}
\label{che01}
\te
and 

\be
0=-\left\langle p^{\lambda}p^{\rho}p^{\nu}\right\rangle\left\{u_{\rho} \dot\beta_{\nu}^{\left(0\right)}
-\Delta^{\mu}_{\rho}\beta_{\nu,\mu}\right\}
\label{che02}
\te
Which we recognize just as the conservation equation for an ideal energy-momentum tensor. Introducing the functions eq. (\ref{Thetanm}),
then

\be 
\left\langle p^{\lambda}p^{\rho}p^{\nu}\right\rangle=t^5\Theta^{0}_{3} \left[u^{\lambda}u^{\rho}u^{\mu}+\frac13\left(u^{\lambda}\Delta^{\nu\rho}+u^{\nu}\Delta^{\rho\lambda}+u^{\rho}\Delta^{\lambda\nu}\right)\right] \label{che03}
\te
Decomposing as usual $\beta_{\mu}=u_{\mu}/t$ we get 

\be
\beta_{\mu,\nu}=u_{\mu}u_{\nu}\frac{\dot t}{t^2}-\Delta^{\lambda}_{\nu}u_{\mu}\frac{t_{,\lambda}}{t^2}+\frac1t\left\{-\Delta^{\lambda}_{\mu}u_{\nu}\dot u_{\lambda}+\frac12\sigma_{\mu\nu}+\frac12\omega_{\mu\nu}+\frac13\Delta_{\mu\nu}u^{\lambda}_{,\lambda}\right\} \label{che04}
\te 
where $\sigma_{\mu\nu}$ is the shear tensor eq. (\ref{shear}) and 

\be
\omega_{\mu\nu}=\left[\Delta_{\mu}^{\rho}\Delta_{\nu}^{\sigma}-
 \Delta_{\nu}^{\rho}\Delta_{\mu}^{\sigma}\right]u_{\rho,\sigma} \label{che05}
\te

\be
\dot\beta_{\mu}=-u_{\mu}\frac{\dot t}{t^2}+\frac1t\dot u_{\mu} \label{che06}
\te
This yields the equations (\ref{che02}) in the familiar form

\begin{eqnarray}
 \frac{\dot t^{ 0}}{t} &=& - \frac{1}{3} u^{\nu}_{,\nu}\label{consEMT0-t}\\
 \dot u^{\mu\left(0\right)} &=& - \Delta^{\mu\nu} \frac{t_{,\nu}}{t} \label{consEMT0-sp}
\end{eqnarray}
We use these equations to eliminate time derivatives from eq. (\ref{che01}), leading to
 
 \begin{equation}
 {Q}\left[t^2 F {Q}\left[\chi_1\right]\right]
 = - \frac{p^{\mu}p^{\rho}}{2t} \sigma_{\mu\rho} \label{1st-order}
\end{equation}
The solution to eq. (\ref{1st-order}) is 

\be 
F {Q}\left[\chi_1\right]=- \frac{p^{\mu}p^{\rho}}{2t^3} \sigma_{\mu\rho} -c_{\mu}p^{\mu}
\label{che07}
\te
We find a new integrability condition

\be 
0=\left\langle \frac{p^{\lambda}p^{\mu}p^{\rho}}F\right\rangle\frac{\sigma_{\mu\rho}}{2t^3}-c_{\mu}\left\langle \frac{p^{\lambda}p^{\mu}}F\right\rangle \label{che08}
\te
The first term is zero and therefore the homogeneous solution vanishes. Now again

\be
\chi_1=- \frac{p^{\mu}p^{\rho}}{2t^3F} \sigma_{\mu\rho} -c'_{\mu}{p^{\mu}}
\label{che09}
\te 
and imposing the constraint eq. (\ref{constraint}) we see that again $c'_{\mu}=0$, whence eq. (\ref{chi-1-2t}).

\subsection{Second order Ch-En solution }\label{gen-1pdf-2order}

We obtain the second order Ch-E solution setting $n=1$ in eqs. (\ref{che3b}) and (\ref{che4b}). Lets start from the integrability condition

\be
0=-u_{\mu} \left\{\left\langle p^{\lambda}p^{\mu}p^{\nu}\right\rangle\dot\beta_{\nu}^{\left(1\right)}+\left\langle p^{\lambda}p^{\mu}\dot\chi_1^{\left(0\right)}\right\rangle\right\}+\Delta^{\mu}_{\rho} \left\langle p^{\lambda}p^{\rho} \chi_{1,\mu}\right\rangle
\label{che10}
\te
Under the linearized approximation

\bea
 \chi_{1,\mu} &=&  - \frac{p^{\rho}p^{\sigma}}{2t^3F} \sigma_{\rho\sigma,\mu} \nn
&=&  - \frac{p^{\rho}p^{\sigma}}{2t^3F} \Lambda_{\rho\sigma}^{\lambda\tau} u_{\lambda,\tau\mu} \label{che11-0}
\tea
in particular

\bea
 \dot\chi^{\left(0\right)}_{1} &=&  - \frac{p^{\rho}p^{\sigma}}{2t^3F} \Lambda_{\rho\sigma}^{\lambda\tau} \dot u^{\left(0\right)}_{\lambda,\tau} \nn
&=&   \frac{p^{\rho}p^{\sigma}}{2t^3F} \Lambda_{\rho\sigma}^{\lambda\tau}  \frac{t_{,\lambda\tau}}{t} \label{che12}
\tea
To evaluate the integrability condition eq. (\ref{che10}) we compute the mean values $\left\langle p^{\lambda}p^{\mu}p^{\nu}\right\rangle$ and $\left\langle p^{\lambda}p^{\mu}p^{\nu}p^{\rho}/F\right\rangle$,
obtaining

\bea
&-&u_{\mu}\left\langle p^{\lambda}p^{\mu}p^{\nu}\right\rangle\dot\beta_{\nu}^{\left(1\right)}=t^4\Theta^0_3\left\{u^{\lambda}\frac{\dot t^{1}}t+\frac13\dot u^{\left(1\right)\lambda}\right\}\nn
&-&u_{\mu}\left\langle p^{\lambda}p^{\mu}\dot\chi_1^{\left(0\right)}\right\rangle=0\nn
&&\Delta^{\mu}_{\rho} \left\langle p^{\lambda}p^{\rho} \chi_{1,\mu}\right\rangle=\frac{-1}{15}t^3\Theta^1_4\sigma^{\lambda\rho}_{,\rho}
\label{che13}
\tea
and we get

\bea
\dot t^{1}&=&0\nn
\dot u^{\left(1\right)\lambda}&=&\frac{1}{5t}\frac{\Theta^1_4}{\Theta^0_3}\sigma^{\lambda\rho}_{,\rho} \label{che14}
\tea

We now turn to eq. (\ref{che3b})

\bea
&&{Q}\left[t^2 F  {Q}\left[\chi_{2}\right]\right]=p^{\mu}u_{\mu} \left\{p^{\nu}\dot\beta_{\nu}^{\left(1\right)}-\dot\chi_1^{\left(0\right)}\right\}-p^{\rho} \Delta^{\mu}_{\rho} \chi_{1,\mu}\nn
&=&p^{\mu}u_{\mu} \left\{p^{\nu}\frac{1}{5t^2}\frac{\Theta^1_4}{\Theta^0_3}\sigma^{\rho}_{\nu,\rho}-\frac{p^{\rho}p^{\nu}}{2t^3F} \Lambda_{\rho\nu}^{\lambda\tau}  \frac{t_{,\lambda\tau}}{t} \right\}+p^{\rho} \Delta^{\lambda}_{\rho}\frac{p^{\nu}p^{\mu}}{2t^3F} \sigma_{\nu\mu,\lambda}
\label{che11}
\tea
A first integration yields

\be
t^2 F  {Q}\left[\chi_{2}\right]=p^{\mu}u_{\mu} \left\{p^{\nu}\frac{1}{5t^2}\frac{\Theta^1_4}{\Theta^0_3}\sigma^{\rho}_{\nu,\rho}-\frac{p^{\rho}p^{\nu}}{2t^3F} \Lambda_{\rho\nu}^{\lambda\tau}  \frac{t_{,\lambda\tau}}{t} \right\}+p^{\rho} \Delta^{\lambda}_{\rho}\frac{p^{\nu}p^{\mu}}{2t^3F} \sigma_{\nu\mu,\lambda}-d_{\mu}p^{\mu} \label{che15}
\te
where the $d_{\mu}$'s are integration constants. 
We thereby find the integrability condition

\be
0=\left\langle \frac{p^{\lambda}}{t^2F}\left\{p^{\mu}u_{\mu} \left[p^{\nu}\frac{1}{5t^2}\frac{\Theta^1_4}{\Theta^0_3}\sigma^{\rho}_{\nu,\rho}-\frac{p^{\rho}p^{\nu}}{2t^3F} \Lambda_{\rho\nu}^{\lambda\tau}  \frac{t_{,\lambda\tau}}{t} \right]+p^{\rho} \Delta^{\tau}_{\rho}\frac{p^{\nu}p^{\mu}}{2t^3F} \sigma_{\nu\mu,\tau}-d_{\mu}p^{\mu}\right\}\right\rangle \label{che16}
\te
which reduces to
\be
0=-\frac{1}{15}\left[\Theta^2_4-\frac{\Theta^1_4}{\Theta^0_3}\Theta^1_3\right]\sigma^{\lambda\rho}_{,\rho}+td_{\mu}\Theta^1_2\left[u^{\lambda}u^{\mu}+\frac13\Delta^{\mu\nu}\right] \label{che17}
\te
or else

\be
d_{\mu}=-\frac1{5t}D_2\sigma^{\rho}_{\mu,\rho} \label{che18}
\te
where $D_2$ was defined in eq. (\ref{dedos})
We thus arrive at the preliminary result

\be
\chi_2=\frac{1}{t^2F}\left\{p^{\mu}u_{\mu} \left[p^{\nu}\frac{1}{5t^2}\frac{\Theta^1_4}{\Theta^0_3}\sigma^{\rho}_{\nu,\rho}-\frac{p^{\rho}p^{\nu}}{2t^3F} \Lambda_{\rho\nu}^{\lambda\tau}  \frac{t_{,\lambda\tau}}{t} \right]+p^{\rho} \Delta^{\tau}_{\rho}\frac{p^{\nu}p^{\mu}}{2t^3F} \sigma_{\nu\mu,\tau}-d_{\mu}p^{\mu}\right\}-d'_{\mu}p^{\mu} \label{che19}
\te 
with a new set of integration constants $d'_{\mu}$. To fix the $d'_{\mu}$ we must enforce the constraint eq. (\ref{constraint}). Introducing the tensor $\Sigma$ from eq. (\ref{derivada-sigma}), we get

\bea 
\chi_2&=&\frac1{t^2}\Sigma_{\nu\mu\rho}\frac{p^{\nu}p^{\mu}p^{\rho}}{2t^3F^2}-\frac1{t^3}u_{\mu}t_{,\lambda\tau}\Lambda_{\rho\nu}^{\lambda\tau}\frac{p^{\nu}p^{\mu}p^{\rho}}{2t^3F^2}\nn
&+&\frac1{5t^3}\sigma^{\rho}_{\nu,\rho}p^{\nu}\left[\frac {D_2}F+\frac{\Theta^1_4}{\Theta^0_3}\frac{p^{\mu}u_{\mu} }{tF}+\left(\frac{p^{\mu}u_{\mu} }{tF}\right)^2\right]-d'_{\mu}p^{\mu}
\label{chidosapp}
\tea
Enforcing the constraint eq. (\ref{constraint}) yields eq. (\ref{chidos}).


\begin{thebibliography}{99}
\bibitem{Rom-Rom-19} P. Romatschke and U. Romatschke, \textit{Relativistic Fluid Dynamics In and Out of Equilibrium And 	Applications to Relativistic Nuclear Collisions}, Cambridge Univ. Press, Cambridge, UK (2019).

\bibitem{calzetta16} E. Calzetta, {Real Relativistic Fluids in Heavy Ion Collisions}, in  \textit{Geometric, Algebraic and Topological Methods for Quantum Field Theory}; L. Cano, A. Cardona,  H. Ocampo, A. F. R. Lega Eds; World Scientific, Singapore, p 155 (2016).

 \bibitem{GraCalKan22} N. Mirón-Granese, E. Calzetta and A Kandus, {Primordial Weibel instability}, JCAP 2022, 028 (2022).

 \bibitem{GraCal18} N. Mirón-Granese and E. Calzetta, {Primordial Gravitational Waves Amplification from Causal Fluids}, Phys. Rev. D {97}, 023517 (2018).

 \bibitem{MiGra21} N. Mirón-Granese, {Relativistic Viscous Effects on the Primordial Gravitational Waves Spectrum} JCAP 2021, 008 (2021).

\bibitem{CalHu08} E. Calzetta and B. L. Hu, \textit{Nonequilibrium Quantum Field Theory}, Cambridge Univ. Press, Cambridge, UK(2008).
 
\bibitem{tanos} L. Rezzolla and O. Zanotti, \textit{Relativistic Hydrodynamics} (Oxford University Press,
Oxford, 2013).

\bibitem{DRlibro}  G. Denicol and D. Rischke,  \textit{Microscopic Foundations of Relativistic Fluid Dynamics} (Springer, Berlin, 2021).

f\bibitem{Isr76} W. Israel, {Nonstationary irreversible thermodynamics: A causal relativistic theory}, Ann. Phys. (NY) {100}, 310 (1976).

\bibitem{LMR86} I. S. Liu, I. M\"uller and T. Ruggeri, 
{Relativistic thermodynamics of gases}, Ann. Phys. {169},
191 (1986).

\bibitem{GL90} R. Geroch and L. Lindblom, {Dissipative relativistic
	fluid theories of divergence type}, Phys. Rev. D {41}, 1855 (1990).

\bibitem{PRCal09} J. Peralta-Ramos and E. Calzetta, {Divergence-type 	nonlinear conformal hydrodynamics}, Phys. Rev. D {80}, 126002 (2009).

\bibitem{BGK54} P. L. Bhatnagar, E. P. Gross, M. Krook. {A Model for Collision Processes in Gases. I. 
Small Amplitude Processes in Charged and Neutral One-Component Systems}, Phys. Rev. 94, 511 (1954)
 







\bibitem{Marle-a} C. Marle, {Sur l'etabissement des équations de l'hydrodynamique des
fluids relativistes dissipatifs. I. - L'équation de Boltzmann relativiste.}, Ann. Inst. Henri
Poincaré (A), 10, 67 (1969).

\bibitem{Marle-b} C. Marle, {Sur l'etabissement des équations de l'hydrodynamique des
fluids relativistes dissipatifs. II. - Méthodes de résolution approchée de l'equation de Boltzmann relativiste.}, Ann. Inst. Henri Poincaré (A), 10, 127 (1969).

\bibitem{AWa} J. L. Anderson and H. R. Witting, {A relativistic Relaxation-Time Model
for the Boltzmann Equation}, Physica 74, 466 (1974)

\bibitem{AWb} J. L. Anderson and H. R. Witting, {Relativistic Quantum Transport Coefficients}, Physica 74, 489 (1974).

\bibitem{DeNo24} G. S. Denicol and J. Noronha, {Spectrum of the Boltzmann collision operator for 
$\lambda\phi^4$ theory in the classical regime}, Phys. Lett. B {850}, 138487 (2024).

\bibitem{RBD23b} G. S. Rocha, C. V. P. de Brito and G. S. Denicol, {Hydrodynamic theories for a system of weakly self-interacting classical ultra-relativistic scalar particles: microscopic derivations and attractors}, Phys. Rev. D {108}, 036017 (2023).

\bibitem{Dud13} M. Dudynski, Spectral properties of the linearized Boltzmann operator in Lp for $1\le p\le\infty$, J. Stat. Phys. 153, 1084 (2013).

\bibitem{LY16} L. Luo and H. Yu, Spectrum analysis of the linearized relativistic Landau equation, J. Stat. Phys. 163, 914 (2016).

\bibitem{Hu24} J. Hu, {Relaxation time approximation revisited and pole/cut structure in retarded 
correlators}, e-print arXiv:2409.05131 (2024).

\bibitem{DMT10} K. Dusling,  G. D. Moore, and D. Teaney, Radiative energy loss and v 2 spectra for viscous hydrodynamics, Phys. Rev. C 81, 034907 (2010). 

\bibitem{LO10} M. Luzum and J.-Y. Ollitrault, Constraining the viscous freeze-out distribution function with data obtained at the BNL Relativistic Heavy Ion Collider (RHIC), Phys. Rev. C 82, 014906 (2010)

\bibitem{RDN21} G. Rocha, G. Denicol and J. Noronha,  Novel Relaxation Time Approximation to the Relativistic Boltzmann Equation , Phys. Rev. Lett. 127, 042301 (2021).



\bibitem{RFDN22} G. Rocha, M. Ferreira,  G. Denicol and J. Noronha,  Transport coefficients of quasiparticle models within a new relaxation time approximation of the Boltzmann equation, Phys. Rev. D 106, 036022 (2022).

\bibitem{PRCal10} E. Calzetta and J. Peralta-Ramos, Linking the hydrodynamic
and kinetic description of a dissipative relativistic conformal theory, Phys. Rev. D 82, 106003 (2010).

\bibitem{PRCal13} J. Peralta-Ramos and E. Calzetta, Macroscopic approximation
to relativistic kinetic theory from a nonlinear closure,
Phys. Rev. D 87, 034003 (2013).

\bibitem{KW19} A. Kurkela and U. Wiedemann, Analytic structure of nonhydrodynamic modes in kinetic theory , Eur. Phys. J. C 79, 776 (2019).

\bibitem{WW21} G. Wilka and Z. Włodarczyk, Beyond the relaxation time approximation, Eur. Phys. J. A57, 221 (2021).

\bibitem{M21} S. Mitra,  Relativistic hydrodynamics with momentum-dependent relaxation time , Phys. Rev. C 103, 014905 (2021).

\bibitem{M22} S. Mitra,  Correspondence between momentum-dependent relaxation time and field redefinition of relativistic hydrodynamic theory, Phys. Rev. C 105, 014902 (2022).

\bibitem{AM23} V. Ambruş and E. Molnár, {Shakhov-type extension of the relaxation time approximation in relativistic kinetic theory and second-order fluid dynamics}, e-print arXiv:2311.11603. 

\bibitem{AM24} V. Ambruş and E. Molnár, {High-order Shakhov-like extension of the relaxation time approximation in relativistic kinetic theory}, e-print arXiv:2401.04017.

\bibitem{H23} J. Hu, Full-order mode analysis within a mutilated relaxation time approximation, eprint arXiv:2310.05606

\bibitem{DBJJ22} D. Dash, S. Bhadury, S. Jaiswal and A. Jaiswal, Extended relaxation time approximation and relativistic dissipative hydrodynamics, Phys. Lett. B, 831, 13720 (2022).

\bibitem{DBJJ23} D. Dash, S. Bhadury, S. Jaiswal and A. Jaiswal,   Relativistic second-order viscous hydrodynamics from kinetic theory with extended relaxation-time approximation,  Phys. Rev. C 108, 064913 (2023).



\bibitem{B23} T. Bhattacharyya, Non-extensive Boltzmann Transport Equation: the Relaxation Time Approximation and Beyond, Physica A 624, 128910 (2023). 

\bibitem{RWDNR24} G. Rocha, D. Wagner, G. Denicol, J. Noronha, and D. Rischke, D.H. Theories of Relativistic Dissipative Fluid Dynamics. Entropy  26, 189 (2024).


\bibitem {S24} M. Strickland, Hydrodynamization and resummed viscous hydrodynamics, arXiv:2402.09571, Contributed chapter to Quark-Gluon Plasma 6 book 


\bibitem{Gav24a} L. Gavassino, Mapping
GENERIC Hydrodynamics into Carter’s Multifluid Theory. Symmetry  16, 78 (2024).

\bibitem{WG24} D. Wagner and L. Gavassino, The regime of applicability of Israel-Stewart hydrodynamics, Phys. Rev. D 109, 016019 (2024).

\bibitem{GaDiNo24a} L. Gavassino, M. M. Disconzi and J. Noronha, {Universality Classes of Relativistic
Fluid Dynamics: Foundations}, Phys. Rev. Lett.  {132}, 222302 (2024).

\bibitem{GaDiNo24b} L. Gavassino, M. M. Disconzi and J. Noronha, {Universality classes of relativistic
fluid dynamics: Applications}, Phys. Rev. D {109}, 096041 (2024).

\bibitem{ChapCow} S. Chapman and T. G. Cowling, \textit{The Mathematical Theory of 
	 Non-Uniform Gases}, 3rd. ed. ,Cambridge Univ. Press, Cambridge (1970).

\bibitem{DNMR12} G. S. Denicol, H. Niemi, E. Moln\'ar and D. H. Rischke, {Derivation of transient relativistic fluid dynamics from the Boltzmann equation}, Phys. Rev. D {85}, 11047 (2012).

\bibitem{WPA22} D. Wagner, A. Palermo and V. Ambrus, {Inverse-Reynolds-dominance approach to 
transient fluid dynamics}, Phys. Rev. D {106}, 016013 (2022).

\bibitem{BKCJ20} S. Bhadury, M. Kurian, V. Chandra and A. Jaiswal, First order dissipative hydrodynamics and viscous corrections to the entropy four-current from an effective covariant kinetic theory, J. Phys. G 47 (2020) 8, 085108

\bibitem{BKCJ21} S. Bhadury, M. Kurian, V. Chandra and A. Jaiswal,  Second order relativistic viscous hydrodynamics within an effective description of hot QCD medium, J. Phys. G:  48,  105104, (2021).

\bibitem{DMMEZ23}S. Diles, A. Miranda, L. Mamani, A. Echemendia and V. Zanchin, Third-order relativistic fluid dynamics at finite density in a general hydrodynamic frame, arXiv:2311.01232.

\bibitem{PeCa21} G. Perna and E. Calzetta, {Linearized dispersion relations in viscous relativistic hydrodynamics}, Phys. Rev. D 104, 096005 (2021).

\bibitem{CanCal20} L. Cantarutti and E. Calzetta, {Dissipative-type theories for Bjorken and Gubser flows}, Int. J. Mod. Phys. A {35}, 2050074 (2020).

\bibitem{AgCal17} M. Aguilar and E. Calzetta, {Causal relativistic hydrodynamics of conformal Fermi-Dirac gases}, Phys, Rev. D {95}, 076022 (2017)

\bibitem{S05} H. Struchtrup, \emph{Macroscopic Transport Equations for Rarefied
Gas Flows} (Springer, Berlin, 2005).


\bibitem{BR79} G. Boillat and T. Ruggeri, {On the evolution law of weak discontinuities
for hyperbolic quasi-linear systems}, Wave Motion {1}, 149 (1979).

\bibitem{BR97a} G. Boillat and T. Ruggeri, {Hyperbolic principal subsystems: entropy 
convexity and subcharacteristic conditions}, Arch. Rational Mech. Anal. {137}, 
305 (1997).

\bibitem{BR97b} G. Boillat and T. Ruggeri, {Moment equations in the kinetic theory
of gases and wave velocities}, Continuum Mech. Thermodyn. {9}, 205
(1997).
\bibitem{BR99} G. Boillat and T. Ruggeri, {Relativistic gas: Moment equations
and maximum wave velocity}, J. Math. Phys. (N.Y.) 40,
6399 (1999).




\bibitem{Mull99} I. M\"uller, {Speeds of propagation in classical and relativistic
extended thermodynamics}, Living Rev. Relativity {2}, 1 (1999).


\bibitem{SM22} S. Mitra, Causality and stability analysis of first-order field redefinition in relativistic hydrodynamics from kinetic theory, Phys. Rev. C 105, 054910 (2022).

\bibitem{HSSW23}  M. Heller, A. Serantes, M. Spaliński and B. Withers, The Hydrohedron: Bootstrapping Relativistic Hydrodynamics, arXiv:2305.07703v2 

\bibitem{ASN23}  N. Abboud, E. Speranza and J. Noronha, Causal and stable first-order chiral hydrodynamics , arXiv:2308.02928.

\bibitem{BRD23}  C. de Brito, G. Rocha and G. Denicol,  Hydrodynamic theories for a system of weakly self-interacting classical ultra-relativistic scalar particles: causality and stability , arXiv:2311.07272.



\bibitem{HK24}  R. Hoult and P. Kovtun, Causality and classical dispersion relations, Phys. Rev. D 109 (2024) 4, 046018. 

\bibitem{WP24}  D-L Wang and S. Pu, Stability and causality criteria in linear mode analysis: stability means causality, Phys. Rev. D 109, L031504 (2024).

\bibitem{GDN24} L. Gavassino, M. Disconzi and J. Noronha, Dispersion relations alone cannot guarantee causality, Phys. Rev. Lett. 132, 162301 (2024).

\bibitem{Isr60} W. Israel, Relativistic theory of shock waves, Proc. R. Soc.
A 259, 129 (1960).

\bibitem{Cal22} E. Calzetta, Steady asymptotic equilibria in conformal relativistic fluids, Phys. Rev. D 105, 036013 (2022).





\bibitem{Isr88} W. Israel, Covariant fluid mechanics and thermodynamics: An introduction,  in A. M. Anile and Y. Choquet-Bruhat (eds.) \textit{Relativistic Fluid Dynamics}(Springer, New York,
1988), p. 152.

\bibitem{Moore24}G. Moore, Hydrodynamics as $v_s\to c$, arXiv:2404.01968v2 [hep-ph] (2024).

\bibitem{Isr72} W. Israel,  The relativistic Boltzmann equation. In: L. O’Raifeartaigh (ed.), \textit{General Relativity: Papers in Honour of J. L. Synge}, pp. 201–241. Clarendon, Oxford (1972).

\bibitem{GLW80} S. R. de Groot, W. A. van Leeuwen, and C. G. van Weert,
\textit{Relativistic Kinetic Theory} (North-Holland, Amsterdam,
1980).

\bibitem{Ste71} J. M. Stewart, \textit{Non-Equilibrium Relativistic Kinetic Theory}
(Springer, New York, 1971).

\bibitem{Lib03} R. Liboff, \textit{Kinetic Theory} (Springer, New York, 2003).

\bibitem{CMK02} C. Cercignani and G. Medeiros Kremer, \textit{The Relativistic
Boltzmann Equation: Theory and Applications} (Birkhauser, Basel, 2002).




\bibitem{Eck40} C. Eckart, \textit{The Thermodynamics of Irreversible
	Processes. III. Relativistic Theory of the Simple Fluid},
Phys. Rev. {58}, 919 (1940)

\bibitem{LL59} L. D. Landau and E. M. Lifshitz, \textit{Fluid
	Mechanics}, Pergamon Press, Oxford (1959).

 
 
\bibitem{DoGaMoShTo22} T. Dore,, L. Gavassino, D. Montenegro, M. Shokri and G. Torrieri,
{Fluctuating relativistic dissipative hydrodynamics as a gauge theory}, Ann. of Phys. 
{442}, 168902 (2022).

\bibitem{BhaMiRoSi24} S. Bhattacharyya, S. Mitra, S. Roy and R. Singh, {Field redefinition and
its impact in relativistic hydrodynamics}, arXiv:2409.15387 [nucl-th] (2024).

\bibitem{BhaMiRo24} S. Bhattacharyya, S. Mitra and S. Roy, {Causality and Stability in relativistic
hydrodynamic theory - a choice to be endured}, Phys. Lett. B {856}, 138918  (2024).

\bibitem{HL83} W. Hiscock and L. Lindblom, Stability and Causality in
Dissipative Relativistic Fluids, Ann. Phys. (NY) 151, 466 (1983).

\bibitem{IS79a} W. Israel and M. Stewart, {Transient relativistic thermodynamics and kinetic theory}, Ann. Phys. (NY) {118}, 341 (1979).










 














\end{thebibliography}
\end{document}